\documentclass[useAMS,usenatbib,referee]{mn2e}

\usepackage{graphicx,amssymb,amsmath}

\def\Msun{M_\odot}
\def\Lsun{L_\odot}
\def\Rsun{R_\odot}
\def\gsun{g_\odot}
\def\FOCP{F_{\rm OCP}}
\def\Fvib{F_{\rm vib}}

\def\Mstar{M_\star}
\def\Lstar{L_\star}
\def\Rstar{R_\star}
\def\gstar{g_\star}
\def\Mdot{\dot{M}}
\def\etareim{\eta_{\rm Reim}}
\def\etasc{\eta_{\rm SC}}

\def\Teff{T_{eff}}

\title[A Stellar Evolution Code for Calculating Complete Tracks]
{A New, Efficient Stellar Evolution Code for Calculating Complete Evolutionary Tracks}

\author[A. Kovetz, O. Yaron and D. Prialnik]
{Attay Kovetz$^{1,2}$\thanks{E-mails: attay@etoile.tau.ac.il (Kovetz); oferya@post.tau.ac.il (Yaron); dina@planet.tau.ac.il (Prialnik)}, 
Ofer Yaron$^{2}$\footnotemark[1] and Dina Prialnik$^{2}$\footnotemark[1]\\
$^{1}$School of Physics and Astronomy, Sackler Faculty of Exact Sciences, Tel Aviv University, Israel\\
$^{2}$Department of Geophysics and Planetary Sciences, Sackler Faculty of Exact Sciences, Tel Aviv University, Israel}

\begin{document}


\pagerange{\pageref{firstpage}--\pageref{lastpage}} \pubyear{2008}

\maketitle

\label{firstpage}

\begin{abstract}
We present a new stellar evolution code and a set of results, demonstrating its capability at calculating
full evolutionary tracks for a wide range of masses and metallicities.
The code is fast and efficient, and is capable of following through all evolutionary phases,
without interruption or human intervention.
It is meant to be used also in the context of modeling the evolution of dense stellar systems,
for performing live calculations for both normal star models and merger-products.

The code is based on a fully implicit, adaptive-grid numerical scheme that solves simultaneously for structure,
mesh and chemical composition. Full details are given for the treatment of convection, equation of state,
opacity, nuclear reactions and mass loss.

Results of evolutionary calculations are shown for
a solar model that matches the characteristics of the present sun to an accuracy of better than 1\%;
a $1\ \Msun$ model for a wide range of metallicities;
a series of models of stellar populations I and II, for the mass range $0.25$ to $64 \Msun$, followed from
pre-main-sequence to a cool white dwarf or core collapse.
An initial final-mass relationship is derived and compared with previous studies.
Finally, we briefly address the evolution of non-canonical configurations, merger-products of low-mass main-sequence
parents.
\end{abstract}


\section{Introduction} \label{intro}

Simulating the evolution of a star requires the solution of a set of partial differential equations with boundary
conditions at the center and surface, involving extensive input physics, such as equations of state, nuclear reactions,
opacities, as well as recipes for treating convection, mass loss, or material mixing. This is accomplished by
complex computer codes that are time consuming and depend on a large number of adjustable parameters, both physical and numerical, 
needed for dealing with evolutionary phases that are different in nature. From the formation to the death of a star, 
differences between evolutionary phases are so large, that studies are usually devoted to---and often codes are devised for---a
specific part of a star's life, ignoring or simplifying, or suppressing others. So far, no code has been suited or applied to obtain
complete, unabridged evolutionary tracks over the entire range of stellar masses and metallicities, although many have come
close to accomplishing this task (e.g. \citealt{1995MNRAS.274..964P}, \citealt{1998MNRAS.298..525P}).  
For example, most (if not all) evolution codes crash at the helium core flash phase.
Most of the stellar evolution codes do not solve simultaneously for the structure and the composition; this introduces
serious errors in some critical phases whenever the mass grid $\{m_i\}$ changes, as it must eventually
\citep{2006MNRAS.370.1817S}.
Our aim has been to develop a versatile and robust stellar evolution code that is free of such handicaps. 

A further demand on the code is efficiency and speed. Furthermore, it should be capable not
only of evolving any star through all phases without intervention, but also of dealing with peculiar objects. 
Such a code could be incorporated into an N-body code that deals with dense stellar systems, if not at present, 
then---given the rapid and continual advance in computing power---in the foreseeable future.
The computation methods of N-body gravitating systems have undergone a revolutionary development owing to the work of 
\citet{1963MNRAS.126..223A} (see review by \citealt{1999PASP..111.1333A}) and gaining impetus in the past 
two decades (e.g. \citealt{2003gmbp.book.....H}, \citealt{2001MNRAS.323..630H}, \citealt{2005MNRAS.363..293H}): not only have new algorithms been developed,
capable of dealing with dense
stellar systems (e.g. \citealt{2001MNRAS.321..199P}, \citealt{2004MNRAS.351..473P}), but also special hardware has been constructed 
under the GRAPE (GRAvity PipE) project \citep{1997ApJ...480..432M}.

However, in order to render these sophisticated N-body calculations realistic, the effect of the structure and
evolution of the constituent stars must be considered as well. This led, less than a decade ago, to the development
of the MODEST (MOdelling DEnse STellar systems) project, whose aim is to combine N-body dynamics with the 
hydrodynamics of stellar collisions on the one hand, and with stellar evolution of the cluster population, 
on the other (see \citet{2003NewA....8..337H}).
So far, studies of stellar systems have resorted to
short cuts based on sets of discrete pre-calculated evolutionary tracks: either interpolating between them, or using
parametrized fit formulae. Clearly, this procedure is incapable of dealing with `non-canonical' stars, the outcome
of collisions and mergers. 

In this paper we thus present a new evolutionary code that we have developed with this aim in mind.
The outline of the code and method of solution are presented in the next section, Section~\ref{code}; the input
physics is described in some detail in Section~\ref{input-phys}, and results of representative calculations are discussed
in Section~\ref{res}.


\section{The Evolution Code} \label{code}

\subsection{Set of Equations and Boundary Conditions} \label{eqs}

The equations that govern the evolution of a star are those of continuity, hydrostatic equilibrium,
energy transfer (radiative or convective), energy balance, and composition balance:

\begin{align}
&\frac{\partial}{\partial m}\frac{4\pi}{3} r^3=\frac{1}{\rho} & , \label{eq:evol1}\\
&\frac{\partial p}{\partial m}=-\frac{Gm}{4\pi r^4} & , \label{eq:evol2}\\
&\frac{\partial \ln T}{\partial m}=\nabla \frac{\partial \ln p}{\partial m} & , \label{eq:evol3}\\
&\frac{\partial u}{\partial t}+p\frac{\partial}{\partial t}\frac{1}{\rho}=q-\frac{\partial L}{\partial m} & , \label{eq:evol4}\\
&F_j=-\sigma_j\frac{\partial Y_j}{\partial m} & , \label{eq:evol5}\\
&\frac{\partial Y_j}{\partial t}=R_j-\frac{\partial F_j}{\partial m} & . \label{eq:evol6}
\end{align}

In these equations, mass $m$ and time $t$ are the independent variables. The dependent ones are radius $r$,
density $\rho$, temperature $T$, and the number fractions $Y_j$, related to the mass fractions $X_j$ 
by $Y_j=X_j/A_j$, where $A_j$ is the $j$'th atomic mass.  
The particle flux $F_j$ of the $j$'th species is assumed to be diffusive (proportional to
the abundance gradient of the $j$'th species), determined by the diffusion coefficient $\sigma_j$.

We regard $(\rho,T,Y)$ as the basic thermodynamic variables. They determine, through the equation of state, the
pressure $p(\rho,T,Y)$ and the specific energy
$u(\rho,T,Y)$, as well as the opacity $\kappa(\rho,T,Y)$, energy production rate $q(\rho,T,Y)$ and nuclear
energy rates $R_j(\rho,T,Y)$ (via
an imported list of tables and formulae). The temperature `gradient' $\nabla(r,L,m,\rho,T,Y)$ and the (convective
mixing) diffusion coefficients $\sigma_j(r,L,m,\rho,T,Y)$ are provided by convection recipes.

The foregoing equations are to be solved subject to the following boundary conditions:
at the centre,
\begin {equation}\label{eq:cbc}
         r=0 ,\qquad  L=0 ,\qquad F_j=0 \quad ;
\end{equation}
at the surface, which we take to be the stellar photosphere,
\begin {equation}\label{eq:sbc}
         \kappa p_G =\left( 1-\Gamma\right) g\tau_s ,\qquad L=4\pi r^2\sigma T^4 ,\qquad F_j=0 \quad .
\end{equation}
In the first member of eq. \eqref{eq:sbc}, $p_G$ is the material (`gas') pressure which, together with the
radiation pressure $p_R$,
makes up the total $p\,$; $\Gamma=\kappa L/4\pi cGm$; $g=Gm/r^2$; and $\tau_s$ is the photospheric optical depth,
which we take to be unity (\citealt{1998ApJ...495..401K}, \citealt{1999PhR...311..383K}).

We shall solve the equations of evolution over a grid of mass points $m_1=0, m_2, \dots, m_n=M$, but we shall follow
(\citealt{1971MNRAS.151..351E}, \citealt{1972MNRAS.156..361E}) in using an {\it adaptive} grid, where the mass points ${m_2, \dots, m_{n-1}}$ depend
on the solution.
Since, by eqs. \eqref{eq:evol1}--\eqref{eq:evol2}, $r$ at the centre varies like $m^{1/3}$, and $p$ like $m^{2/3}$,
we replace $m$ by $x=m^{2/3}$, and $r$ by $s=r^2$, in this pair of equations. Equations \eqref{eq:evol1}--\eqref{eq:evol6}
then become
{\allowdisplaybreaks
\begin{subequations} \label{eq:setgrp}
\begin{align}
        &ds=\frac{3}{4\pi\rho}\left(\frac{x}{s}\right)^{\frac{1}{2}}dx & , \label{eq:set1}\\
        &d\ln p=-\frac{3G}{8\pi p}\left(\frac{x}{s}\right)^2dx & , \label{eq:set2}\\
        &d\ln T=\nabla d\ln p & , \label{eq:set3}\\
        &dL=\left[ q-\frac{\delta u+p\delta\frac{1}{\rho}}{\delta t}\right]dm & , \label{eq:set4}\\
        &F_j=-\sigma_j\frac{dY_j}{dm} & , \label{eq:set5}\\
        &dF_j=\left( R_j-\frac{\delta Y_j}{\delta t}\right)dm & . \label{eq:set6}
\end{align}
\end{subequations}
}
These may be regarded as differential equations, written in terms of differentials; alternatively, they may be
thought of as representing difference equations.
In the latter case, at the centre, the indeterminate ratio $x/s=0/0$ is replaced by its limit $(4\pi\rho_1)/3)^{2/3}$,
where $\rho_1$ is the central value of the density. The change from $r$ to $s$ obviously requires
appropriate changes (such as $s=0, g=Gm/s$) in the boundary conditions.

The equations of structure and composition are solved simultaneously with a mass distribution function, implementing
an adaptive mesh. This is done by requiring constant increments of a monotonic function of the form
\begin{equation} \label{eq:meshf}
        f=(m/M)^{2/3}+c_1X_H-c_2\ln p-c_3\ln\frac{T}{T+c_4} ,
\end{equation}
where the $c$'s are appropriate non-negative constants. Near the centre the requirement of equal increments of
$f$ will lead to equal increments of $x=m^{2/3}$. The second term of $f$ will force equal increments of the
hydrogen mass fraction
where $X_H$ changes rapidly (at an H-burning shell). The third will lead to equal steps of $\ln p$ towards the
surface, where $m/M\approx 1$ and $X_H$ is uniform; and the last term will cause a fine subdivision around
$T=c_4\approx 20,000 K$, where the opacity varies rapidly over several orders of magnitude.

\subsection{Numerical Scheme} \label{num}

The variables $(s,m,L,\rho,T,p,Y_j)$ are represented by arrays over a grid of $i=1,\dots,n$, where
$i=1$ corresponds to the centre, and $i=n$ to the surface (photosphere). Thus eqs. \eqref{eq:set1}--\eqref{eq:set2}
become the difference equations
\begin{align}
       &s_i-s_{i-1}=\frac{1}{2}\left[\frac{3}{4\pi\rho_{i-1}}\left(\frac{x_{i-1}}{s_{i-1}}\right)^{1/2}
        +\frac{3}{4\pi\rho_i}\left(\frac{x_i}{s_i}\right)^{1/2}\right](x_i-x_{i-1}) & , \label{eq:speq1}\\
       &\ln p_i-\ln p_{i-1}=-\frac{1}{2}\left[\frac{3G}{8\pi p_{i-1}}\left(\frac{x_{i-1}}{s_{i-1}}\right)^2
        +\frac{3G}{8\pi p_i}\left(\frac{x_i}{s_i}\right)^2\right](x_i-x_{i-1}) & . \label{eq:speq2}
\end{align}
There is one such pair of equations for each $i=2,\dots,n$. Together with the boundary conditions $s_1=0$ and
$(\kappa p_G)_n=(1-\Gamma_n)g_n\tau_s$, these add up to $2n$ equations.

The variables $\nabla$, $L$ and $F_j$, related to the energy and particle fluxes,
are replaced by arrays that refer to the {\it midpoints} $i\pm 1/2$. Thus eqs. \eqref{eq:set3}--\eqref{eq:set6} become
\begin{align}
       &\ln T_i-\ln T_{i-1}=\nabla_{i-\frac{1}{2}}(\ln p_i-\ln p_{i-1}) & , \label{eq:TLYeq1}\\
       &L_{i+\frac{1}{2}}-L_{i-\frac{1}{2}}
        =\left[ q_i-\frac{\delta u_i}{\delta t}-p_i\frac{\delta\frac{1}{\rho_i}}{\delta t}\right]
          \frac{1}{2}(m_{i+1}-m_{i-1}) & , \label{eq:TLYeq2}\\
       &F_{i+\frac{1}{2}}=-\sigma_{i+\frac{1}{2}}\frac{Y_{i+1}-Y_i}{m_{i+1}-m_i} & , \label{eq:TLYeq3}\\
       &F_{i+\frac{1}{2}}-F_{i-\frac{1}{2}}
        =\left( R_i-\frac{\delta Y_i}{\delta t}\right) \frac{1}{2}(m_{i+1}-m_{i-1}) & . \label{eq:TLYeq4}
\end{align}
where, in the last pair of equations, we have suppressed the index $j$ that refers to the nuclear species.
The coefficients $\nabla_{i-\frac{1}{2}}$ and $\sigma_{i+\frac{1}{2}}$ are evaluated by using the arithmetic
means of the grid-point arguments, for example $r_{i-\frac{1}{2}}=(r_{i-1}+r_i)/2$.  
Again, there is one eq. \eqref{eq:TLYeq1} for each $i=2,\dots,n$, which, together with the boundary condition
$L_n=4\pi r_n^2\sigma T_n^4$, brings the number of equations up to $3n$. Furthermore, there is one set of
eqs. \eqref{eq:TLYeq2}
and \eqref{eq:TLYeq4} for each $i=1,\dots,n$ (and for each one of the species). If $J$ is the number of species,
the number of equations becomes $(4+J)n$.
At $i=1$ we set $L_{i-\frac{1}{2}}=F_{i-\frac{1}{2}}=0$ and $m_{i-1}=0$ in eqs. \eqref{eq:TLYeq2}
and \eqref{eq:TLYeq4}, which takes
care of the central boundary conditions $L=F=m=0$. At $i=n$ we set $L_{i+\frac{1}{2}}=L_n$,
$F_{i+\frac{1}{2}}=0$ and $m_{i+1}=m_n$ in eqs. \eqref{eq:TLYeq2} and \eqref{eq:TLYeq4}. This is in accord
with the surface boundary conditions.

The requirement of equal increments of the mesh function $f$ is simply
\begin{equation} \label{eq:meq}
        f_{i+1}-f_i=f_i-f_{i-1} .
\end{equation}
There is one such equation for each $i=2,\dots,n-1$. At the ends $i=1$ and $i=n$ we respectively impose the two
boundary conditions
\begin{equation} \label{eq:mbc}
        m_1=0,\qquad m_n=M+\dot M\delta t ,
\end{equation}
where $\dot M(m_n,r_n,L_n)$ is the rate of mass accretion---or loss, if negative. Thus
we have a total of $(5+J)n$ equations for $(5+J)n$ variables---5 arrays $(s,L,m,\rho,T)$ and $J$ arrays $Y_j$,
each array being of length $n$.

The partial time derivatives, $\partial u(m,t)/\partial t$, etc., have been replaced, respectively, by difference ratios
$\delta u/\delta t$, etc. When a configuration at a previous time is available, $\delta u$ is usually taken to
be $u(m,t)-u(m,t-\delta t)$, where $u(m,t)$ is iterated upon. The solution 
of eqs. \eqref{eq:set1}--\eqref{eq:set6} then has the accuracy $O(\delta t)$.
It should be noted
that $u(t-\delta t)$ is represented by a grid function over a (previous) set of $m_i$'s that will not generally
include the
$m$ for which $u(m,t-\delta t)$ is desired. We therefore determine $u(m,t-\delta t)$ by interpolation, using cubic
Hermite splines. These splines have the advantage that, if the grid function vanishes at two consecutive
$m_i$'s, the interpolant will not dip below zero anywhere between them. This is especially important when interpolating
the number fractions $Y_j$.

Except at the first time step, the previous, as well as the anteprevious, configurations are available. Instead of a
chord through $u(m,t)$ and $u(m,t-\delta t)$, we can then pass a {\it parabola} through $u(m,t)$, $u(m,t-\delta t)$
and $u(m,t-\delta t-\delta t')$, and evaluate its derivative at $t$. If this derivative is again denoted
by $\delta u/\delta t$, we have
\begin{align}
\delta u=\alpha u(m,t)+\beta u(m,t-\delta t)+\gamma u(m,t-\delta t-\delta t'),  \label{eq:parab1}
\end{align}
where
\begin{equation}
\alpha=\frac{\delta t'+2\delta t}{\delta t'+\delta t},\qquad \beta=-\frac{\delta t'+\delta t}{\delta t'}
,\qquad \gamma=\frac{(\delta t)^2}{(\delta t'+\delta t)\delta t'} .\label{eq:parab2}
\end{equation}
This leads to a solution with accuracy $O(\delta t^2)$. Of course $u(m,t-\delta t-\delta t')$, like
$u(m,t-\delta t)$, has to be determined by interpolation.

The $(5+J)n$ nonlinear eqs. \eqref{eq:speq1}--\eqref{eq:mbc} for the the arrays $(s,L,m,\rho,T,Y_j)$
are solved simultaneously by Newton-Raphson iterations. This requires, at each iteration stage, the solution of a linear
system with a band matrix of order $(5+J)n$, and band width $15+4J$.

The derivatives required by the Newton-Raphson method are evaluated analytically whenever possible. In the
case of opacities, which are obtained from tables with the aid of cubic Hermite spline interpolation, we
use the (analytic) derivatives of the splines. Numerical derivatives are only used for the energy generation
and loss rates, because the neutrino loss rates are provided by cumbersome fit formulae.

\subsection{Computational Details}  

Our automatically varying timesteps, determined mainly by limits imposed on
the maximal changes (a few percent), and on the number of Newton-Raphson iterations,
allowed during a timestep, span a wide dynamic range---from seconds/minutes during core or shell
flashes to several times $10^8$ or even $10^9$ years in the main-sequence phase (of low-mass stars).
With a relative accuracy of $\sim$0.0001, the typical number of Newton-Raphson iterations is 3--4.
The grid mass shells, determined by the mass-distribution function, span a range of $\sim 10^{-15} \Msun$
(in a WD atmosphere) to $\gtrsim 10^{-1}\ \Msun$ (in an inert stellar core). There is an option of fixing the mass
grid, which we are forced to use during the WD cooling phase, when the mass array $\{m_i\}$ ceases to
be monotonically increasing in double precision arithmetic.
With these features in mind, the typical number of grid points may be as low as 150 or 200; a typical number of
timesteps for a complete evolutionary track is 1000; and typical execution time is of the order of $10\ (\pm 5)$ minutes
on a portable computer (Pentium 4 and higher). The latter is, however, strongly dependent on both physical behaviour (e.g. mass-loss
rate or the amount of evolutionary phases taking place) and computational prescriptions (required outputs/interfaces).

The code---targeted for Unix/Linux machines---is written in Fortran 90 and 
consists of an online graphical interface using Tim Pearson's PGPLOT.


\section{Input Physics} \label{input-phys}

\subsection{Equation of State (EOS)} \label{eos}

The EOS is derived from a free energy, which is a sum of ionic, radiative and electronic contributions,
together with corrections for pressure ionization, Coulomb interactions and quantum effects:

\begin{align} \label{eq:freeen}
      F=&\sum_i F(T,V,N_i)-\frac{1}{3}aT^4V  \notag\\
        &+\Omega(T,V,\mu_F)+(\mu_F-mc^2)N \notag\\
        &+F_{PI}+F_{CQ} .
\end{align}
where $V$ is the volume, $a$ is Stefan-Boltzmann's constant, $\mu_F$ is the Fermi chemical potential, 
and $m$ is the electron's mass.
The free energy of the $N_i$ particles of the $i$'th ionic species is
\begin{align} \label{eq:freeion}
     &F(T,V,N_i)=(kT\ln\frac{N_i}{z_i}-kT+\chi_i)N_i     , \notag\\
     &z_i=VQ_i\ell_i^{-3},\qquad \ell_i=\frac{h}{\sqrt{2\pi m_ikT}},
\end{align}
where $\chi_i$ is the reference energy (relative to the completely ionized state) of the $i$'th
ion, and $Q_i$ is its partition function. The thermal length $\ell_i$ depends on the temperature
and on the $i$'th particle's mass $m_i=A_im_H$, where $A_i$ is the atomic or molecular weight.

We take account of ionization equilibria for hydrogen and helium; heavier elements (the `metals')
are assumed to be completely ionized. In the stellar envelope, where the metals amount to at most
a few percent by mass, and a few thousandths by number, this introduces an error that is much
smaller than other uncertainties in the EOS.
In a carbon/oxygen stellar core, the metals are pressure-ionized in any case. Ionization equilibria
of the metals play an important role in determining the opacity, but we use opacity tables that
are entirely independent of our EOS.

Remembering that the reference energies for the completely ionized species H$^+$ and He$^{++}$
are zero by definition, the $\chi_i$ for H, H$_2$, H$^+$, He, He$^+$ and He$^{++}$ are, respectively,
-13.598, -31.673, 0, -79.003, -54.416 and 0 (in eV). Also, $\chi_i=0$ for the metals.

Except for the case of H$_2$, we replace the partition function $Q_i$ by a constant statistical weight
$g_i$, which is 1 for H$^+$, He, He$^{++}$ and all metals, and 2 for H and He$^+$. For the hydrogen
molecule, we use our own table of $Q_{H_2}(T)$, which we have calculated, using the molecular constants
of \citet{1966PDAO...13....1T}; see also \citet{1987A&A...182..348I}.

Electrons and positrons are described by the fermion grand thermodynamic potential
\begin{align} \label{eq:fermpot}
        \Omega(T,V,\mu_F)= - Cmc^2V\int_\beta^\infty
                          \Gamma(\epsilon/\beta)D_+(\epsilon,\phi)\,d\epsilon/\beta ,
\end{align}
where
\begin{align} \label{eq:fermpot1}
        &C=\frac{1}{\pi^2}\left(\frac{mc}{\hbar}\right)^3, \qquad \Gamma(x)=\frac{1}{3}(x^2-1)^\frac{3}{2} , \notag\\
        &D_\pm(\epsilon,\phi)=\frac{1}{e^{\epsilon-\phi}+1}\pm\frac{1}{e^{\epsilon+\phi}+1} , \notag\\
        &\beta=\frac{mc^2}{kT}, \qquad \phi=\frac{\mu_F}{kT},
\end{align}
\citep{1967ApJ...150..131R}.
The Fermi chemical potential $\mu_F$, which includes the rest-mass energy, is connected with the number
{\it difference}, electrons minus positrons, through
\begin{align} \label{eq:Neq}
       N&=-\Omega_{\mu_F}(T,V,\mu_F) \notag\\
       &=CV\int_\beta^\infty\Gamma'(\epsilon/\beta)D_-(\epsilon,\phi)\,d\epsilon/\beta = N_e-N_p ,
\end{align}
where $\Omega_{\mu_F}(T,V,\mu_F)$ denotes the partial derivative $\partial\Omega(T,V,\mu_F)/\partial\mu_F$.
Clearly the positron contribution, which is due to the second term of $D_\pm$,
becomes insignificant whenever $\phi$ is large (say $\phi\geq15$).
The last equation determines $\mu_F(T,V,N)$ as a function of $T$, $V$ and $N$ (actually the Fermi parameter
$\epsilon_F=(\mu_F-mc^2)/kT$ in terms of $T$ and $N/V$). The number difference $N$ must satisfy
the equation of charge neutrality
\begin{equation} \label{eq:neutrality}
      N=\sum Z_iN_i.
\end{equation}

If, in the expression \eqref{eq:freeen} for the free energy, electrons appeared only in the second line, then it
would follow that
\begin{align}
     F_N(T,V,N)&=\Omega_{\mu_F}\mu_{F,N}+\mu_{F,N}N+\mu_F-mc^2 \notag\\
               &=\mu_F-mc^2 ,
\end{align}
where the subscript $N$ denotes the partial derivative with respect to $N$, at constant $T$ and $V$.
Thus $\mu_F$ would indeed be the electron chemical potential $\mu=F_N+mc^2$. We maintain the distinction
(between $\mu$ and $\mu_F$) because other parts of the free energy---for example the pressure ionization
term $F_{PI}$---too, depend on the electron number density.

The pressure $p=-F_V$, the entropy $S=-F_T$  
and their derivatives require derivatives of
$\Omega$, with respect to $\beta$ or $\phi$, up to the second order.
This leads to five additional Fermi-Dirac integrals, in which $\Gamma(\epsilon/\beta)$ is replaced
by $\Gamma'(\epsilon/\beta)$,
$(\epsilon/\beta)\Gamma'(\epsilon/\beta)$,
$\Gamma''(\epsilon/\beta)$, $(\epsilon/\beta)\Gamma''(\epsilon/\beta)$ or
$(\epsilon/\beta)^2\Gamma''(\epsilon/\beta)$. In the degenerate case, when $\epsilon_F=\phi-\beta>5$, $\Omega$
is calculated by Sommerfeld's method, and then differentiated.
Otherwise the six integrals involving the first, electronic, part of $D_\pm$ are calculated in one swoop, using Gaussian
quadrature. The nodes and weights for this quadrature are calculated at the beginning of the run, and their number
can be chosen by the user (the code's default is 12 nodes). The positronic contribution, which is due to the
second part of $D_\pm$, is then obtained by using the same procedure,
with $\phi$ replaced by $-\phi$. The last step is only carried out when $\phi<15$; otherwise, positrons
are ignored.

The pressure ionization term in the free energy is taken from \citet{1995MNRAS.274..964P}:
\begin{align}
    &F_{PI}=-N_ekTg(n_e,T)+N_{e0}kTg(n_{e0},T),  \notag\\
    &g(n_e,T)=e^{-(c_1/x)^{c_2}}[y+\epsilon_F+c_3\ln(1+x/c_4)], \notag\\
    &x=n_em_H,\quad n_e=N_e/V,\quad y=13.60/kT,  \notag\\
    &(c_1,c_2,c_3,c_4)=(3,0.25,2,0.03),
\end{align}
where the 13.60 is in eV, and the units of $c_1$ and $c_4$ are {\rm g cm}$^{-3}$.
Furthermore, $N_{e0}$ is the total number of electrons, bound or free, and $n_{e0}=N_{e0}/V$.
The object of $F_{PI}$ is to induce pressure ionization by reducing the electronic chemical potential
as the number of electrons $n_ea_0^3$ in a cube with side $a_0$, the Bohr radius, increases.
Of course $F_{PI}$ tends to zero as ionization becomes complete, that is, as $N_e\rightarrow N_{e0}$.

The last term, $F_{CQ}$, in the free energy depends on the Coulomb parameter
$\Gamma$ and on the Debye parameter $\Lambda$. For a one-component plasma (OCP)
\begin{equation} \label{eq:Coulomb_par_i}
      \Gamma_i=\frac{Z_i^2e^2}{r_ikT},
\end{equation}
where $Z_i$ is the atomic number, $e$ is the electron charge and $r_i=(4\pi N_i/3V)^{-1/3}$
is the ion-sphere radius. For a mixture, we replace this by
\begin{equation} \label{eq:Coulomb_par}
      \Gamma=\frac{\sum X_iZ_i^2/A_i}{\sum X_iZ_i/A_i}
              \Big[ \Bigl( \frac{\sum X_iZ_i/A_i}{\sum X_i/A_i}\Bigr) ^2 \frac{4\pi n_e}{3}\Bigr]^{1/3}\frac{e^2}{kT},
\end{equation}
where the sums refer to a fully ionized plasma mixture with mass fractions $X_i$. Again, for a
one component plasma,
\begin{equation} \label{eq:Debey_par_i}
      \Lambda_i=\frac{\hbar\omega_{pi}}{kT},\qquad \omega_{pi}^2=\frac{4\pi Z_i^2e^2n_i}{m_i},
\end{equation}
where $\omega_{pi}$ is the plasma frequency. For a mixture, we replace this by
\begin{equation} \label{eq:Debey_par}
      \Lambda=\frac{\hbar\omega_p}{kT},\qquad \omega_p^2=\Bigl(\sum X_iZ_i/A_i\Bigr)N_A4\pi e^2n_e,
\end{equation}
where $N_A$ is Avogadro's number.
The expression for $\FOCP$ takes different forms for the gas-liquid and for the solid phases, and
is based on the work of \citet{1992ApJ...388..521I}. Noting that the OCP form of the translational part
(that is, setting $Q_i=1$ and omitting the $\chi_i$'s) of the ionic free energy
$\sum F(T,V,N_I)$ is
\begin{align}
      \frac{F^0}{\sum N_ikT}&=3\ln\Lambda-1.5\ln\Gamma+\frac{1}{2}\ln\frac{\pi}{6}-1 \notag\\
                            &=3\ln\Lambda-1.5\ln\Gamma-1.32351,
\end{align}
\citet{1992ApJ...388..521I} write the OCP free energies in the form
\begin{align} \label{eq:FOCP}
    \Big( \frac{\FOCP}{\sum N_ikT} \Bigr)_{liq}&=3\ln\Lambda-1.5\ln\Gamma-1.32351-H_l(\Gamma)+J(\Lambda), \\
    \Big( \frac{\FOCP}{\sum N_ikT} \Bigr)_{sol}&=\frac{\Fvib}{\sum N_ikT}-H_s(\Gamma),
\end{align}
where $J(\Lambda)$ takes care of the quantum effects in the gas-liquid phase,
\begin{align} \label{eq:H_liqsol}
      H_l=&\frac{\sqrt{3}}{3}\Gamma^{3/2}+\Gamma^3(-0.104584 \notag\\
         &    +0.172110\ln\Gamma-0.033724\Gamma^{3/2}),&\quad &\Gamma\le1, \notag\\
      H_l=&0.897744\Gamma-3.801720\Gamma^{1/4}+0.758240\Gamma^{-1/4}   \notag\\
         & +0.814871\ln\Gamma +2.584778,&\quad &1\le\Gamma\le200,  \notag\\
      H_s=&0.895929\Gamma+\frac{1612.5}{\Gamma^2},
\end{align}
and $\Fvib(\Lambda)$ is the vibrational contribution to the free energy
\citep{1970A&A.....8..398K}. \citet{1992ApJ...388..521I} have shown that $\Fvib/\sum N_ikT$
can be fitted by a weighted sum of two Debye free energies:
\begin{align} \label{eq:Fvib}
      \frac{\Fvib}{\sum N_ikT}=\alpha L\bigl(\frac{\Lambda}{\Lambda_1}\bigr)
                             +(1-\alpha)L\bigl(\frac{\Lambda}{\Lambda_2}\bigr),
\end{align}
where $\alpha=0.5711$, $\Lambda_1=1.0643$, $\Lambda_2=2.9438$, and $L(x)$ is given by
\begin{align} \label{eq:Deb}
     L(x)=\frac{9}{8}x+3\ln(1-e^{-x})-D(x),\qquad D(x)=\frac{3}{x^3}\int_0^x\frac{t^3dt}{e^t-1}.
\end{align}
The function $J(\Lambda)$ is known \citep{1972A&A....16...72S} to have the the high-temperature limit $\Lambda^2/12$.
At low temperatures the OCP liquid should resemble a bcc lattice, with the ions vibrating about their
equilibrium positions. This leads to a $J(\Lambda)$ proportional to $\Lambda$: according to
\citet{1992ApJ...388..521I}, $J(\Lambda)\rightarrow1.06980\Lambda$ (although their foregoing
fit for $\Fvib$ yields $\Fvib\rightarrow0.76758\Lambda$). They then suggest a functional form
for $J(\Lambda)$ that interpolates between these limits. But this leads to a non-monotonic entropy
(T-derivative of the gas-liquid $\FOCP$); in particular, the specific heat has the required $T^3$ dependence at
low $T$, but with the wrong sign!

Rather than adopt Iben et al.'s $J(\Lambda)$, we note
that, for $\Lambda << 1$, $\Fvib/\sum N_ikT$ tends to $3\ln\Lambda-1-1.49602$, whereas
for $\Lambda >> 1$ it tends to $0.76758\Lambda$, and therefore set
\begin{equation} \label{eq:FOCPliq}
      \Big( \frac{\FOCP}{\sum N_ikT} \Bigr)_{liq}=\frac{\Fvib}{\sum N_ikT}+1.49602-0.32351-1.5\ln\Gamma-H_l(\Gamma).
\end{equation}
The OCP free energies include the contribution of the translational degrees of freedom. Since our
free energy already includes $\sum F(T,V,N_i)$, we must, in order to obtain $F_{CQ}$, {\it subtract}
$F^0$ from each one of the $\FOCP$'s. Thus, finally,
\begin{align} \label{ eq:F_CQ }
      \Big( \frac{F_{CQ}}{\sum N_ikT} \Big)_{liq}&=\frac{\Fvib}{\sum N_ikT}-3\ln\Lambda+2.49602-H_l(\Gamma), \notag\\
      \Big( \frac{F_{CQ}}{\sum N_ikT} \Big)_{sol}&=\frac{\Fvib}{\sum N_ikT}-3\ln\Lambda+1.32351+1.5\ln\Gamma-H_s(\Gamma).
\end{align}
Formally, the difference between the liquid and solid free energies leads to a phase transition when
\begin{equation} \label{eq:phasetr}
     1.49602-H_l(\Gamma)=0.32351+1.5\ln\Gamma-H_s(\Gamma) .
\end{equation}
The root of this equation, the `melting $\Gamma$', is $\Gamma_m=178.2119\,$. We avoid this complication
by interpolating for $F_{CQ}$ in the interval $(\Gamma_m-2,\Gamma_m+2)$.

We shall not pause to write down the equations---such as $\mu_{H_2}=2\mu_H$ for H$_2\leftrightarrow2$H,
or $\mu_{He}=\mu_{He^+}+\mu-mc^2$ for He$\,\leftrightarrow\,$He$^+ +$e---that determine the various states
of hydrogen or helium (e.g. \citet{1995MNRAS.274..964P}).

\subsection{Opacities} \label{opac}

The opacities, which generally depend on density, temperature and composition, are of two kinds:
radiative and conductive. For the radiative part we use Boothroyd's
interpolation program\footnote{ Website http://www.cita.utoronto.ca/$\sim$boothroy/kappa.html .} to interpolate within the
OPAL Rosseland mean opacity tables \citep{1996ApJ...464..943I}. Each one of the OPAL tables is for
a given hydrogen mass fraction $X$, a given total heavy element mass fraction $Z$ (distributed in
accordance with one of a number of standard `mixes'), a given carbon
mass fraction excess $X_C$ (such that the total carbon mass fraction is $X_C\,$, plus the carbon mass fraction
contained in $Z$), and a given oxygen mass fraction excess $X_O\,$. The helium mass fraction is of course
$1-X-Z-X_C-X_O\,$.

Each one of the OPAL tables spans a temperature range $3.75<\log T<8.70$ and a range $-8<\log R<+1$ of $\log R$
values, where $R=\rho/T_6^3\,$, with a cutout at the high $T$, high $R$ corner, and sometimes at the low $T$,
low $R$, corner.
Boothroyd's interpolation program
provides the OPAL opacity $\kappa$, together with its density and temperature derivatives.
(In this section, $T$ is in degrees Kelvin, $\rho$ in ${\rm gr}\,{\rm cm}^{-3}\,$, and $\kappa$ in
${\rm cm}^2\,{\rm g}^{-1}\,$.)

At the low temperature end the OPAL opacities are supplemented by the
\citet{2005ApJ...623..585F} tables.
These span a temperature range $2.70<\log T<4.50$, and the same $R$ range as the OPAL tables. But their
$Z$ range has the upper limit $Z=0.10$, and there is no provision for C or O excesses. We interpolate
among them with a $Z$ value equal to the lesser of $Z+X_C+X_O$ and 0.1.

At the high temperature end, $\log T>8.70$, we extend the OPAL opacities by using electron/positron
scattering opacity according to the fit of \citet{1975ApJ...196..525I}:
\begin{equation} \label{eq:kes}
   \kappa_{es}=[0.2-D-(D^2+0.0004)^{1/2}]2n_{ep}/(N_A\rho),\ D=0.05(\log T_6-1.7) ,
\end{equation}
where $n_{ep}$ is the {\it sum} of the electron and positron number densities.

Electronic conductivities are taken from the \citet{2007ApJ...661.1094C}
tables. These span the temperature range
$3 < \log T < 9\ K$, and the density range $-6 < \log\rho < 9.75\ $. There is one such table for each
value of the atomic number $Z_{ion}$, in fact 15 tables spanning the range $1 < Z_{ion} < 60$. We
use the interpolation program provided by \citet{2007ApJ...661.1094C},
with $Z_{ion}$ equal to the square root of the average (by number) squared atomic number
$$(\sum Z_i^2X_i/A_i) / (\sum X_i/A_i) .$$
The conductivity is converted to a conductive opacity and---harmonically---combined with the radiative opacity.

The various opacity interpolation programs provide the opacity $\kappa$, together with its density
and temperature derivatives. But an evolution code that simultaneously solves for the stellar structure
{\it and} composition requires the derivatives of $\kappa$ with respect to composition as well. One way
to get these is to evaluate the opacity at neighbouring compositions and then form difference ratios.

\begin{figure}
\begin{center}
\includegraphics[scale=.8]{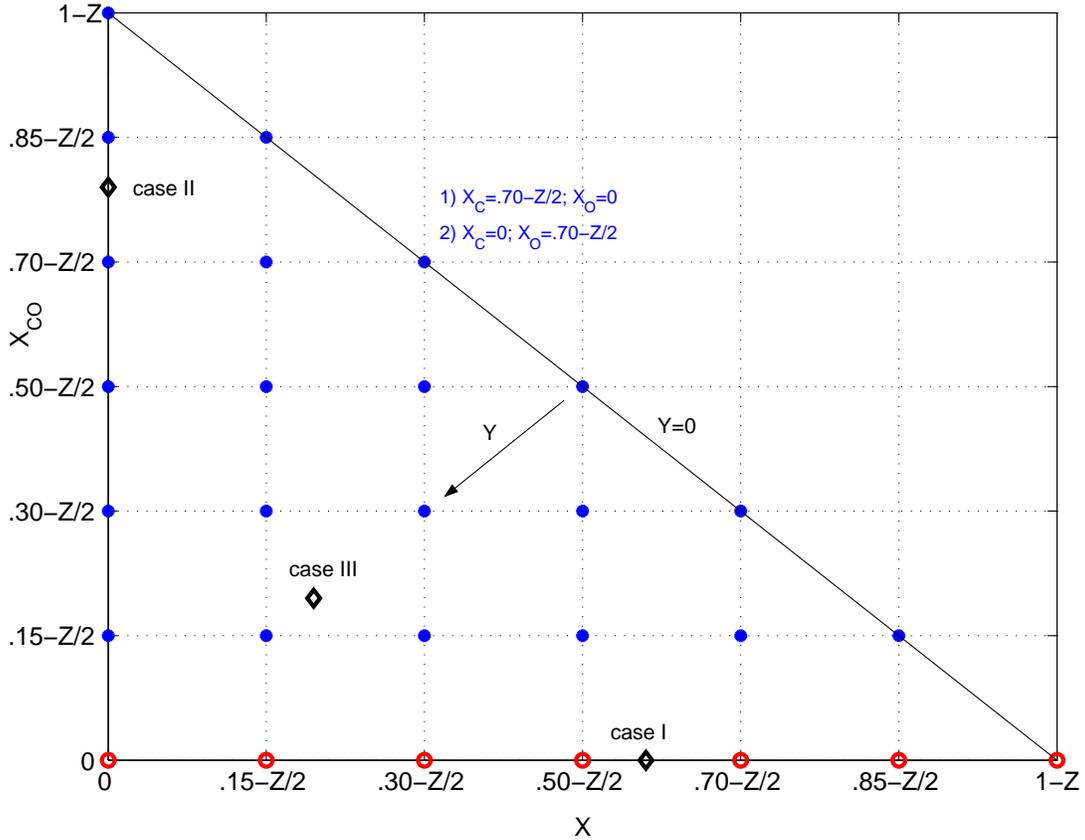}
\end{center}
\caption{
A Schematic representation of our set of 49 opacity tables---spanning a triangular shape in $[X,X_{CO}]$ space---in
between which interpolations are performed for a given metallicity Z.
The 7 open circles along the x-axis denote the 7 tables for zero CO excesses.
Each point within the remaining 21 dots represents two tables:
the excess being completely in carbon for one and completely in oxygen for the other
(such as noted as example for the $(X=.30-Z/2,\ X_{CO}=.70-Z/2)$ position).
The hypotenuse of the triangle relates to zero helium mass fraction $(Y=0)$.
}
\label{fig:opac}
\end{figure}

Alternatively, we use the following method, which yields continuous opacity derivatives:
at the beginning of the evolutionary run, we use the various
interpolation programs to create a set of total---radiative and conductive---opacity tables that,
for the initial stellar model's $Z$, span the
triangular region of Fig.~\ref{fig:opac}. Along the $x$-axis of this figure we have seven values of the hydrogen
mass fraction $X$, from 0 to $1-Z\,$, {\it with no carbon or oxygen mass excesses}. Along the $y$-axis
there are seven values of the {\it combined} C/O excess $X_{CO}=X_C+X_O\,$, again from 0 to $1-Z\,$.
Each point with positive $X_{CO}$ corresponds to a pair of tables: one with carbon excess equal $X_{CO}$ and
zero oxygen excess (that is, `excess all carbon'), and the other one with the same total excess $X_{CO}$,
but 'excess all oxygen'. The $X_{CO}>0$ tables have the lower limit $\log T=4.00$, because the low-temperature
Ferguson-Alexander tables correspond to zero C/O excesses.

There are no points to the right of the hypotenuse (because they would correspond to negative helium
mass fraction $1-Z-X-X_{CO}\,$). Our total number of opacity tables is 49, and these replace the
much larger number of OPAL, Ferguson-Alexander, and Cassisi tables. During MS hydrogen burning,
the stellar core follows a path---from right to left---along the $x$-axis. 
During the HB (Horizontal Branch, core helium burning), C/O excesses rise and the core follows
an upward path along the $y$-axis. Convective mixing may require the evaluation of the opacity in material
containing both hydrogen and C/O excesses, that is, at points inside the triangle of Fig.~\ref{fig:opac}.

In interpolating within the set of tables represented in Fig.~\ref{fig:opac}, we distinguish between three cases:

Case I---No CO excesses. Interpolation is performed only within the 7 $x$-axis (hydrogen) tables.
Within each table we use cubic Hermite splines to interpolate in $\log R$ and $\log T$, in order to
obtain $\log\kappa$ and its $T$ and $R$ derivatives.
Among the seven resulting values of $\log\kappa$, we then interpolate in order to obtain the final opacity
value, together with its $X$ derivative, for the required hydrogen mass fraction. Similar interpolations
among the seven $T$ derivatives, and among the seven $R$ derivatives,
yield the $T$ and $R$ derivatives for the required $X\,$. (Since $R=\rho/T_6^3\,$, the $\rho$ derivative is
simply related to the $T$ and $R$ derivatives.)

Case II---no hydrogen---interpolation within y-axis tables
(where for each $X_{CO}$ value there are two tables, the excess being completely in C for one, and completely in O for the second).
We begin as in case I, by interpolating first among the `excess all carbon' tables, and then among the `excess all oxygen' tables. 
The final value of $\log\kappa$ is then obtained by linear interpolation:
\begin{equation} \label{eq:lin}
   \log\kappa=\frac{X_C}{X_{CO}}\log\kappa_C+\frac{X_O}{X_{CO}}\log\kappa_O ,
\end{equation}
$\log\kappa_C,\ \log\kappa_O$ denoting the opacities as obtained separately from tables
for which excesses are all in C and and from tables for which excesses are all in O, respectively.
Composition derivatives of the opacity, with respect to $X_C$, or with respect to $X_O$, are then obtained
from the last formula. The final $T$ and $R$ derivatives are obtained by similar, linear interpolations.

In comparing $\log\kappa$ obtained by this method with the one returned by Boothroyd's interpolation
(which has its own uncertainties), we found deviations of no more than a few percent. And the largest
of these were at fairly low temperatures, $\sim 5.5 \lesssim \log T \lesssim 6.5$,
where CO-rich opacities are less likely to be needed.

Case III---both C/O excess and hydrogen---interpolations inside the triangle of Fig.~\ref{fig:opac}.
This is a combination of Cases I and II.

\begin{figure}
\centering
\scalebox{0.4}
{\includegraphics{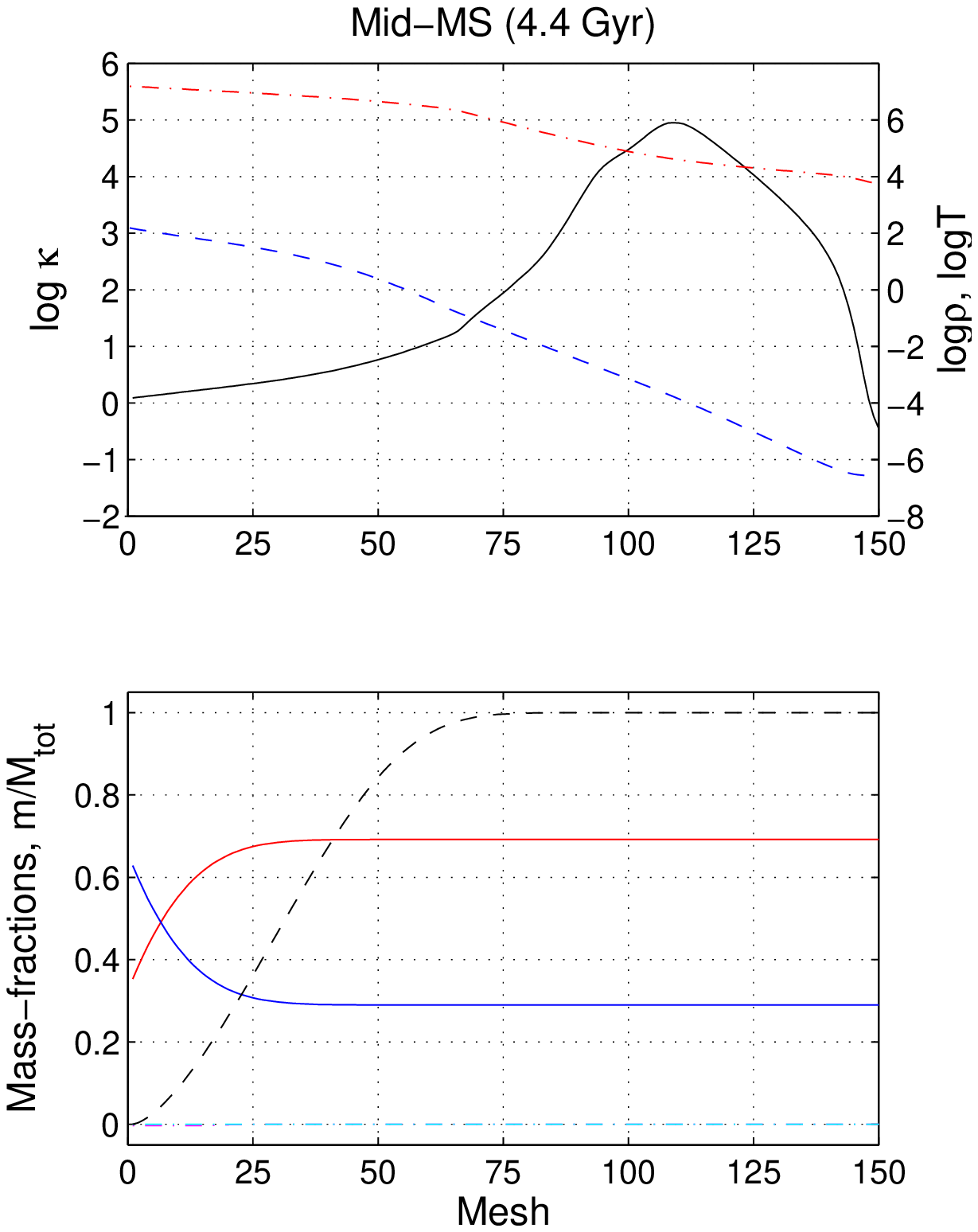}
\includegraphics{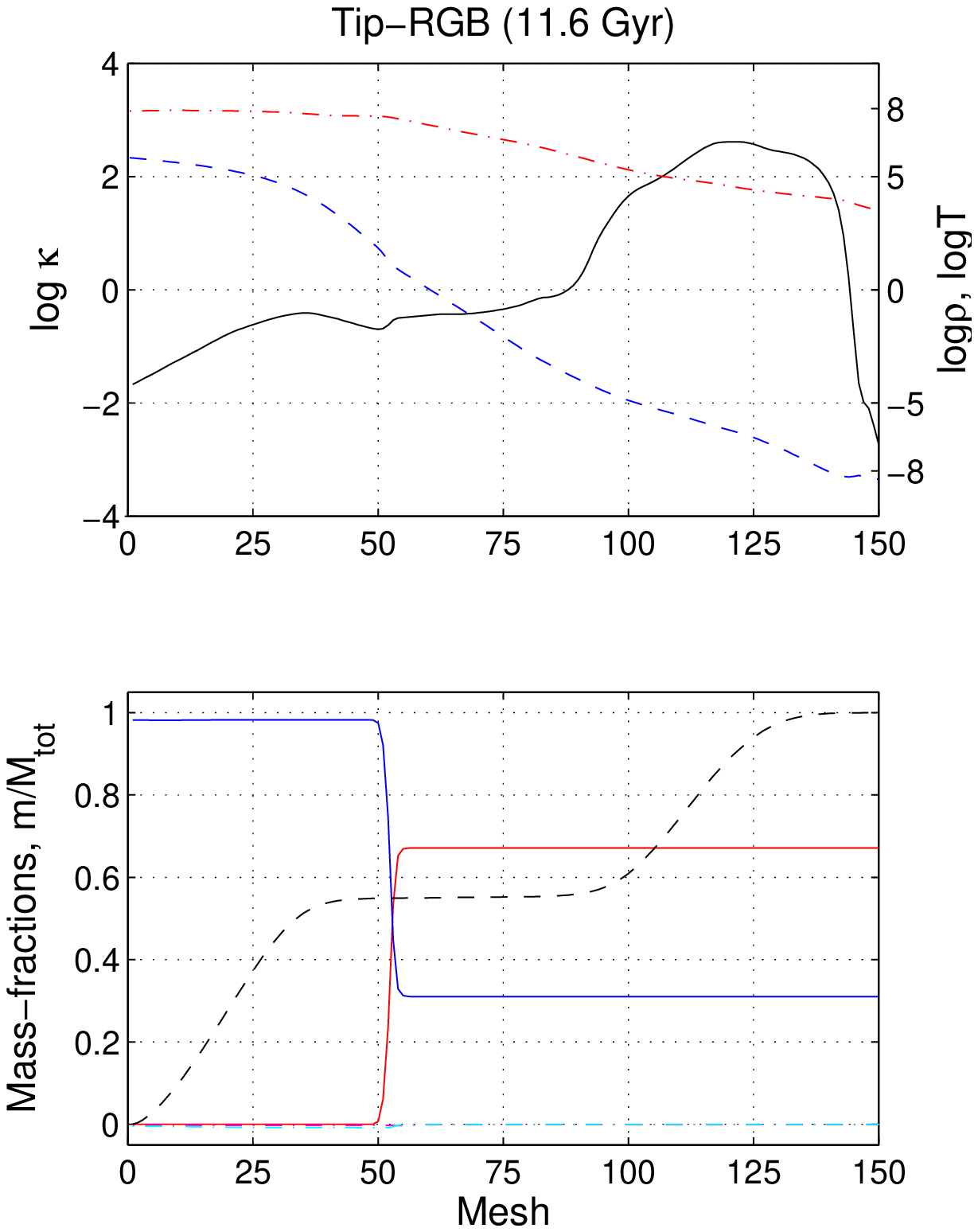}
\includegraphics{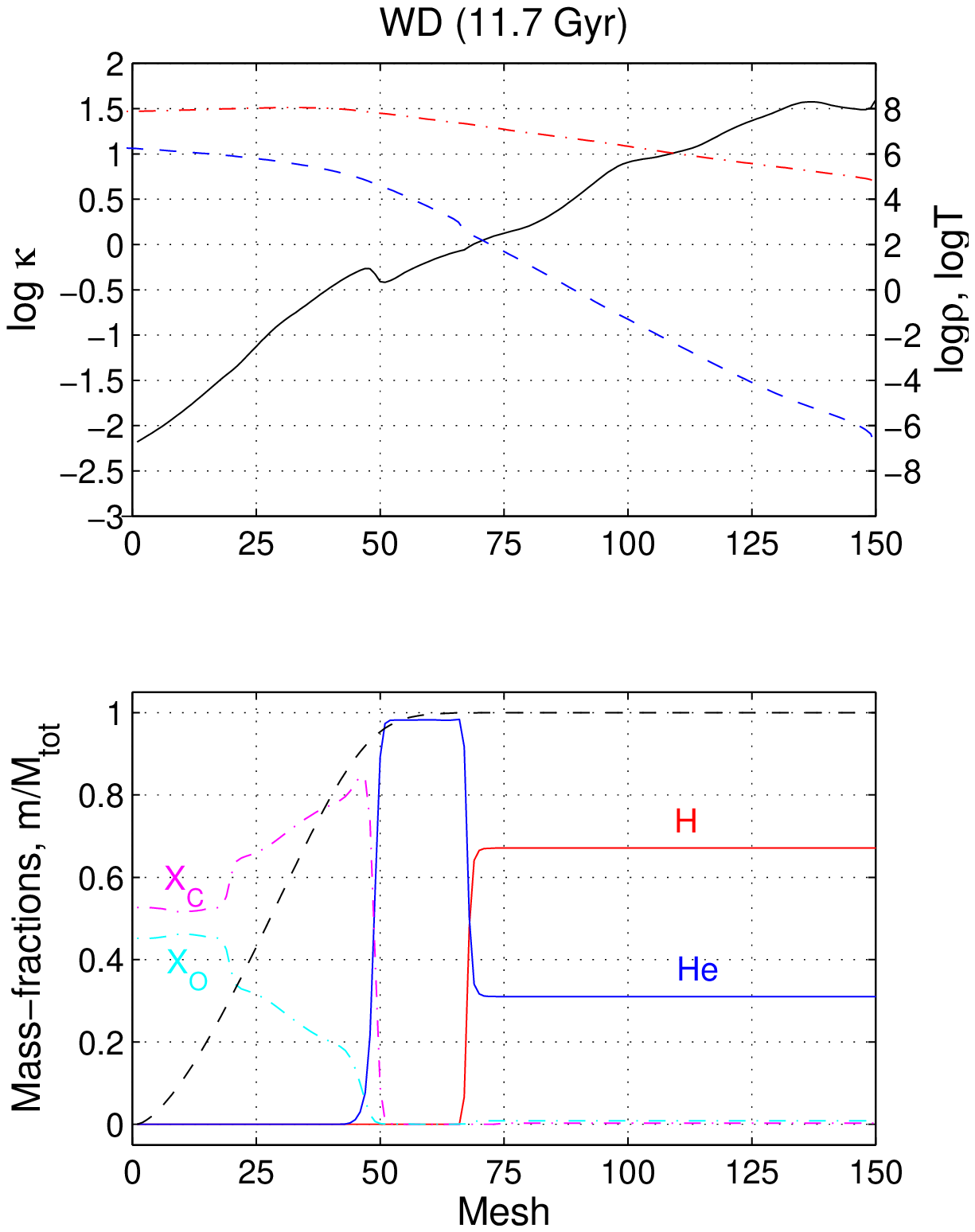}}
\caption{Profiles of internal structure at three snapshots during evolution of a solar model - Mid-MS (left), tip of RGB (middle) and cooling WD (right).
Top panels display profiles of opacity (solid black), density (dashed blue) and temperature (dot-dahsed red).
Bottom panels display internal composition in terms of elemental mass fractions - hydrogen (solid red), helium (solid blue),
and excesses of carbon and oxygen (dot-dashed magenta and cyan, respectively;
values for the excesses are representing closely those of total C and O mass fractions).
Dashed black plots display $m/M_{tot}$; $M_{tot}$ equal $1.00,\ 0.81$ and $0.55\ \Msun$ for the three profiles, left to right, respectively.
Only in the rightmost panel are the carbon and oxygen excesses non-zero (post core helium burning).}
\label{fig:opac_prof}
\end{figure}

In Fig.~\ref{fig:opac_prof} we display opacity profiles, alongside temperature, density and composition profiles,
at three snapshots during a solar model evolution (\S \ref{solar})---Mid-MS, tip of RGB and the end-state as a cooling WD.
Note that ranges of the y-axes values differ in between the three snapshots (columns),
and it is apparent that there is a general decrease in opacity with the advance of evolution.
The peak of opacity at low temperatures (around a few $10^4$ K), close to the surface, is due to the ionization of hydrogen.

\subsection{Nuclear Reaction Rates} \label{nuc}

We use the following nuclear reaction network:
\\

\begin{tabular}{llll}
 & $^1$H(p,$\beta^+\nu$)$^2$D(p,$\gamma$)$^3$He                           &\qquad\qquad & $^4$He($\alpha$)$^8$Be$^*$($\alpha,\gamma$)$^{12}$C \\
 & $^3$He($^3$He,2p)$^4$He                                                &\qquad\qquad & $^{12}$C($\alpha,\gamma$)$^{16}$O \\
 & $^3$He($^4$He,$\gamma$)$^7$Be                                          &\qquad\qquad & $^{14}$N($\alpha,\gamma$)$^{18}$F($\frac{1}{2}\alpha,\gamma$)$^{20}
$Ne \\
 & $^7$Be(e$^-,\nu$)$^7$Li(p,$\alpha$)$^4$He                              &\qquad\qquad & $^{16}$O($\alpha,\gamma$)$^{20}$Ne    \\
 & $^7$Be(p,$\gamma$)$^8$B($\beta^+\nu$)$^8$Be$^*$($\alpha$)$^4$He        &\qquad\qquad & $^{20}$Ne($\alpha,\gamma$)$^{24}$Mg   \\
 &                                                                        &\qquad\qquad & $^{12}$C($^{12}$C,$\alpha$)$^{20}$Ne  \\
 & $^{12}$C(p,$\gamma$)$^{13}$N($\beta^+\nu$)$^{13}$C(p,$\gamma$)$^{14}$N &\qquad\qquad & $^{16}$O($^{16}$O,$\gamma$)$^{32}$S($\gamma,\alpha$)$^{28}$Si \\
 & $^{14}$N(p,$\gamma$)$^{15}$O($\beta^+\nu$)$^{15}$N(p,$\alpha$)$^{12}$C &\qquad\qquad & $^{24}$Mg($\alpha,\gamma$)$^{28}$Si   \\
 & $^{14}$N(p,$\gamma$)$^{15}$O($\beta^+\nu$)$^{15}$N(p,$\gamma$)$^{16}$O &\qquad\qquad & $^{20}$Ne($\gamma,\alpha$)$^{16}$O    \\
 & $^{16}$O(p,$\gamma$)$^{17}$F($\beta^+\nu$)$^{17}$O(p,$\alpha$)$^{14}$N &\qquad\qquad & $^{24}$Mg($\gamma,\alpha$)$^{20}$Ne
\end{tabular}
\\

\noindent
with rates taken from \citet{1988ADNDT..40..283C}\footnote{ Website http://www.phy.ornl.gov/astrophysics/data/cf88/ .}. 
The enhancement of the nuclear reactions by electron screening is taken into account by following the
prescriptions of \citet{1973ApJ...181..457G}.

Where several reactions are written in a chain, the later reactions are taken to be in transient equilibrium
with the first one.
The first five reactions---which constitute the pp-chain---are also assumed to be in transient equilibrium
with each other, so that only the two major isotopes, $^1$H and $^4$He, need to be followed.
Similarly, in the next four reaction chains---which constitute the CNO
cycle---only the major isotopes $^{12}$C, $^{14}$N, and $^{16}$O are followed, and all other
isotopes are taken to be in transient equilibrium.

The triple-alpha reaction $^4$He($\alpha$)$^8$Be$^*$($\alpha,\gamma$)$^{12}$C, together with the four
following lines, constitute helium burning, which involves two further major isotopes---$^{20}$Ne and $^{24}$Mg.
The reaction $^{18}$F($\frac{1}{2}\alpha,\gamma$)$^{20}$Ne
is of course a fiction \citep{1995MNRAS.274..964P}, intended to avoid the creation of $^{22}$Ne,
which is thus replaced by $^{20}$Ne.

Carbon burning proceeds---with comparable probabilities---through the two main branches 
$^{12}$C($^{12}$C,p)$^{23}$Na , $^{12}$C($^{12}$C,$\alpha$)$^{20}$Ne. Since the protons released by the first
one interact with other species, in particular through the reaction $^{23}$Na(p,$\alpha$)$^{20}$Ne, the net result
of carbon burning can be described by the single reaction $^{12}$C($^{12}$C,$\alpha$)$^{20}$Ne \citep{2007nps..book.....I}.

Oxygen burning proceeds via many branches: the main product is $^{28}$Si, with $^{32}$S a close second ({\it ibid.}).
We take $^{28}$Si as our last major isotope. Thus, after oxygen burning, our $^{28}$Si mass fraction is
actually the sum of $X(^{28}$Si) and $X(^{32}$S).

Carbon burning $^{12}$C($^{12}$C,$\alpha$)$^{20}$Ne, neon photodisintegration $^{20}$Ne($\gamma,\alpha$)$^{16}$O,
and oxygen burning $^{16}$O($^{16}$O,$\gamma$)$^{32}$S($\gamma,\alpha$)$^{28}$Si, 
all release $\alpha$ particles that
can be captured by $^{16}$O, $^{20}$Ne, or $^{24}$Mg through the reactions listed above.

In accordance with the foregoing remarks, we need only follow changes in eight active isotopes, namely
$^1$H, $^4$He, $^{12}$C, $^{14}$N, $^{16}$O, $^{20}$Ne, $^{24}$Mg, and $^{28}$Si.
Thus, in eqs. \eqref{eq:evol5}--\eqref{eq:evol6}, or in eqs. \eqref{eq:set5}--\eqref{eq:set6}, the index
$j$ runs over the active isotopes, from 1 to 8. [The number of active isotopes may be changed, provided
that the nuclear reaction network is modified accordingly.] Other isotopes, such as $^{40}$Ca or $^{56}$Fe, are regarded
as inert: they contribute to the EOS, but their abundances do not change; in particular, they do not
undergo convective mixing. For consistency, then, their abundances should be uniform throughout the
initial stellar configuration, and so they will remain.

\subsection{Neutrino Losses} \label{ntloss}

Neutrino losses are according to \citet{1996ApJS..102..411I}, accounting for neutrino formation processes
of pair annihilation, photo annihilation, plasma decay, bremsstrahlung and (optionally) recombination.
There is also an option for using the older fitting formulae of \citet{1967ApJ...150..979B}.

\subsection{Convection (diffusive-convective mixing)} \label{convec}

So long as the radiative 'temperature gradient' $d\ln T/d\ln p\,$, defined by
\begin{equation} \label{eq:nabr}
\nabla_{\rm R}=\frac{\kappa L}{4\pi cGm}\frac{p}{4p_{\rm R}} ,
\end{equation}
does not exceed the adiabatic
\begin{equation} \label{eq:naba}
\nabla_{\rm A}=\frac{\partial\ln T(p,s.X)}{\partial\ln p} ,
\end{equation}
the actual gradient $\nabla\,$, which appears in eq. \eqref{eq:set3}, is equal to $\nabla_{\rm R}\,$,
and the diffusion coefficients $\sigma_j$ in eq. \eqref{eq:set5} are all zero: there is neither convective heat
transport, nor any convective mixing. We hope that the specific entropy $s$ in \eqref{eq:naba} will not be
confused with $s=r^2$.

Convection is taken to set in whenever
\begin{equation} \label{eq:convdelnab}
\Delta\nabla=\nabla_{\rm R}-\nabla_{\rm A}>0   .
\end{equation}
The actual gradient $\nabla\,$ is then calculated in accordance with the
mixing length recipe \citep{1978stat.book.....M}:
consider the non-dimensional (inverse) convective efficiency parameter
\begin{equation} \label{eq:convb}
b=\frac{16\sqrt{2}\sigma_{\rm R}T^4}{(\ell/H)\sqrt{Q}\rho c_{\rm P}Tv_0\tau_e} ,
\end{equation}
where $\sigma_{\rm R}$ is Stefan's constant;
$\ell$ is the mixing length, which we take to be a constant multiple (of order unity) of the pressure
scale height $H=v_0^2/g\,$, where $v_0^2=p/\rho$ is the squared thermal speed and $g=Gm/r^2$ is the
local acceleration of gravity; $Q=-\partial\ln\rho(p,T,X)/\partial\ln T\,$;
$c_{\rm P}$ is the specific heat at constant pressure; and
$\tau_e=\kappa\rho\ell\,$. Let $x$ be the root of the cubic equation
\begin{equation} \label{eq:cubic}
\frac{3}{4b'}x^3 +x^2+2b'x=1 ,
\end{equation}
where
\begin{equation} \label{eq:convbp}
b'=b/\sqrt{\Delta\nabla} ,
\end{equation}
Then the actual gradient is given by
\begin{equation} \label{eq:nabla}
\nabla=\nabla_{\rm A}+(x^2+2b'x)\Delta\nabla .
\end{equation}
It is readily seen that $\nabla\rightarrow\nabla_{\rm A}$ as $b'\rightarrow 0\,$, and $\nabla\rightarrow\nabla_{\rm R}$
as $b'\rightarrow\infty\,$.

Convective mixing is taken to be due to diffusion in a gas of particles---representing the convective
elements---moving at the convective speed
\begin{equation} \label{eq:convv}
v_c=\frac{\ell}{H}\Bigl[\frac{Q}{8}\Delta\nabla\Bigr]^{1/2}xv_0
\end{equation}
\citep{1978stat.book.....M}, with the mean free path $\ell\,$. In such a gas the diffusion coefficient is $\sim v_c\ell\,$.
But in eqs. \eqref{eq:set5}--\eqref{eq:set6} the derivatives are with respect to mass, not radius. We
therefore set the convective diffusion coefficients equal to
\begin{equation} \label{eq:convdif}
\sigma_j=\Bigl(\frac{dm}{dr}\Bigr)^2v_c\ell=(4\pi r^2\rho)^2v_c\ell ,
\end{equation}
the same for all species $j\,$.

The code sometimes runs into difficulties with the foregoing convective diffusion coefficients. We therefore
retain an option whereby the last formula is replaced by a much simpler one:
\begin{equation} \label{eq:convdif_schem}
\sigma_j=k_c\Bigl(\frac{\Delta\nabla}{\nabla_{\rm R}}\Bigr)^2 ,
\end{equation}
where $k_c < 1$ is a numerical coefficient. Its purpose is to ensure that convective mixing does not occur
too suddenly. The value of $k_c$ is related to the evolutionary time scale, and ranges from $\sim0.1$ for
low-mass stars to $\sim0.001$ for massive ones.

Finally, the code has an option for introducing convective overshoot. This is done in a rather schematic
way: at each iteration, after determining the convective zones in accordance with the inequality $\Delta\nabla>0\,$,
we repeat the determination of the zone boundaries, this time with
$\Delta\nabla=\nabla_{\rm R}-\nabla_{\rm A}+\nabla_{\rm OS}>0\,$, where
$\nabla_{\rm OS}$ is a small, positive constant. The temperature gradient $\nabla$ and the convective
diffusion coefficient are then determined by the foregoing formulae, but with the new, augmented, $\Delta\nabla\,$.
We do not attempt to fix $\nabla_{\rm OS}$ by any dependence on local conditions \citep{1995MNRAS.274..964P}.

\subsection{Mass Loss} \label{ml}

The stellar mass may change with time at a prescribed rate $\dot M$, according to boundary condition \eqref{eq:mbc}.
This rate is generally taken to be a function of the stellar parameters $\Mstar, \Lstar, \Rstar\,$.
Over the years, several formulae have been suggested in the literature, each fitting observations of stars in a particular
evolutionary phase. We mention them briefly below, with the mass loss rate (MLR) in units of $\Msun$~yr$^{-1}\,$.

1. The earliest such expression is Reimers's formula \citep{1975MSRSL...8..369R}, derived from observations of RGB stars,
\begin{equation} \label{eq:ml-rei75}
     \Mdot_{\rm Reim}=-4\times 10^{-13}\etareim \frac{\Lstar \Rstar}{\Mstar}  ,
\end{equation}
where the coefficient $\etareim$ lies between $0.3$ and $3.0$.

2. A fit for early type O and B stars, with somewhat modified powers of $\Mstar, \Lstar, \Rstar\,$, is given by \citet{1981ApJ...245..593L}:
\begin{equation} \label{eq:ml-lam}
     \Mdot_{\rm Lam}=-10^{-4.83} \Big(\frac{\Lstar}{10^3}\Bigr)^{1.42} \Big(\frac{\Rstar}{30}\Bigr)^{0.61} \Big(\frac{\Mstar}{30}\Bigr)^{-0.99}  .
\end{equation}

3. A modification of Reimers's MLR, allowing for a superwind on the AGB, is given by \citet{1983A&A...127...73B}:
\begin{equation} \label{eq:ml-bh83}
     \Mdot_{\rm BH}=\Mdot_{\rm Reim}\times \frac{M_{env,0}}{M_{env}}  ,
\end{equation}
where $M_{env,0}$ is the envelope mass at the base of the AGB.

4. Another variation on Reimers's MLR, similar to $\Mdot_{\rm Lam}$, is given by \citet{1990A&A...231..134N} 
(subsequent paper to \citealt{1988A&AS...72..259D}, where $\Mdot$ was given as a function of $\Teff,L$):
\begin{equation} \label{eq:ml-ndj}
     \Mdot_{\rm NDJ}=-9.63\times 10^{-15} \Lstar^{1.42} \Rstar^{0.81} \Mstar^{0.16}  .
\end{equation}

5. The strong increase in mass-loss rate during the AGB stage is rendered by the MLR formulae of \citet{1995A&A...297..727B},
which are based on an investigation of long period variables and shock-driven winds by \citet{1988ApJ...329..299B}.
Bl{\"o}cker's MLR formula is:
\begin{equation} \label{eq:ml-b1}
     \Mdot_{\rm B1}=4.83\times 10^{-9} \Mdot_{Rei} M_{ZAMS}^{-2.1} \Lstar^{2.7}  ,
\end{equation}
and a variant, $\Mdot_{\rm B2}$, has $M_{ZAMS}$ replaced by $\Mstar$.

6. Yet another modification of Reimers's formula, intended for cool winds that are not driven by molecules or dust, is given by
\citet{2005ApJ...630L..73S}, \citet{2007A&A...465..593S}:
\begin{equation} \label{eq:ml-sc05}
     \Mdot_{\rm SC}=-\etasc \frac{\Lstar \Rstar}{\Mstar}\Big(\frac{\Teff}{4000K}\Bigr)^{3.5}
                  \Big(1+\frac{\gsun}{4300\gstar}\Bigr)  ,
\end{equation}
with $\etasc= 8(\pm 1)\times 10^{-14}\, $.
Here two new factors are included, taking into account the dependence of chromospheric height on surface gravity
and the dependence of the mechanical energy flux on the effective temperature.

In applying any of the MLR expressions, instead of turning it on suddenly,
we multiply it by a Fermi weight function
\begin{equation} \label{eq:ml-Fermi}
F(\Rstar)=\frac{1}{1+e^{(R_{thresh}-\Rstar)/(0.05R_{thresh})}}  ,
\end{equation}
where $R_{thresh}$ is an MLR threshold radius, which we typically choose between 1 and 50.
Its precise value is not important, so long as the MLR is negligible for $R=R_{thresh}$.
As $\Rstar$ increases, $F(\Rstar)$ varies smoothly near $R_{thresh}$ from 0 to 1,
over a width of $0.05R_{thresh}$.
This prevents an on-off situation, which can ruin the convergence of the iteration process by which
the difference equations of \S\ref{num} are solved.

The question remains, which formula to use? The code includes an algorithm that identifies the evolutionary
stage of the stellar model by testing various parameters (such as luminosity, radius, composition profiles) and their rates
of change. Therefore, one may pass --- in a smooth manner --- from one formula to another. In this work, we used
\eqref{eq:ml-rei75} for the RGB and \eqref{eq:ml-b1} for later stages. The parameter $\etareim$ was taken progressively higher
with increasing initial mass. The effect of $R_{thresh}$ and $\etareim$ on the results will be briefly discussed in section \S \ref{ifmr}. 

\section{Evolution Sequences} \label{res}

Using the evolution code described in the previous section,
we performed calculations over a wide range of initial stellar masses and metallicities.
In the following sections we address representative results, outcome of continuous
calculations that yield complete evolutionary tracks, starting from an initial pre-main-sequence state
and ending with either a cooling white dwarf (for initial masses below $9~\Msun$), or core collapse of
a configuration resembling a supernova progenitor (for higher initial masses). We use the following
acronyms: MS - main sequence; ZAMS - sero-age main sequence; pre-MS - pre-main-sequence; RGB - red giant branch;
HeF - helium flash; HB - horizontal branch;
AGB - asymptotic giant branch; TP - thermal pulse; WD - white dwarf; HRD - Hertzsprung-Russell diagram.
Central properties are denoted by subscript $c$.

\subsection{Solar Model} \label{solar}

We started from a 'pre-MS' configuration of $1\,\Msun$, of uniform composition $Y=0.29$ and $Z=0.018$---the latter
with a heavy element distribution according to \citet{1993oee..conf...14G}---and a radius of $2.7\,\Rsun$. With a
mixing-length to scaleheight ratio $\alpha\equiv {l/H_P}=2.5$, this configuration reached the ZAMS after
$0.05\ {\rm Gyr}$. At an age of $4.60\ {\rm Gyr}$---which includes the $0.05\ {\rm Gyr}$ from pre-MS to
ZAMS---the model reached a radius of $1.006\ \Rsun$, a luminosity of $1.009\ \Lsun$, and central
characteristics 
$\ T_c=15.59\times10^6\,{\rm K},
\ p_c=2.453\times10^{17}\,{\rm dyn}\,{\rm cm}^{-2}, \ \rho_c=157.9\,{\rm gr}\,{\rm cm}^{-3}$.
We regard this as a good match to the present sun, and the central characteristics in agreement with 
those obtained by other codes (e.g., \citealt{1995MNRAS.274..899R}, \citealt{1998ApJ...504..539T}, \citealt{2000A&A...353..771M}). 
It should, perhaps, be noted that our mixing-length recipe
uses the constants of \citet{1978stat.book.....M}, and our choice of $\alpha=2.5$ may correspond to
different values for other choices of the constants. 

\begin{figure}
\begin{center}
\includegraphics[scale=.8]{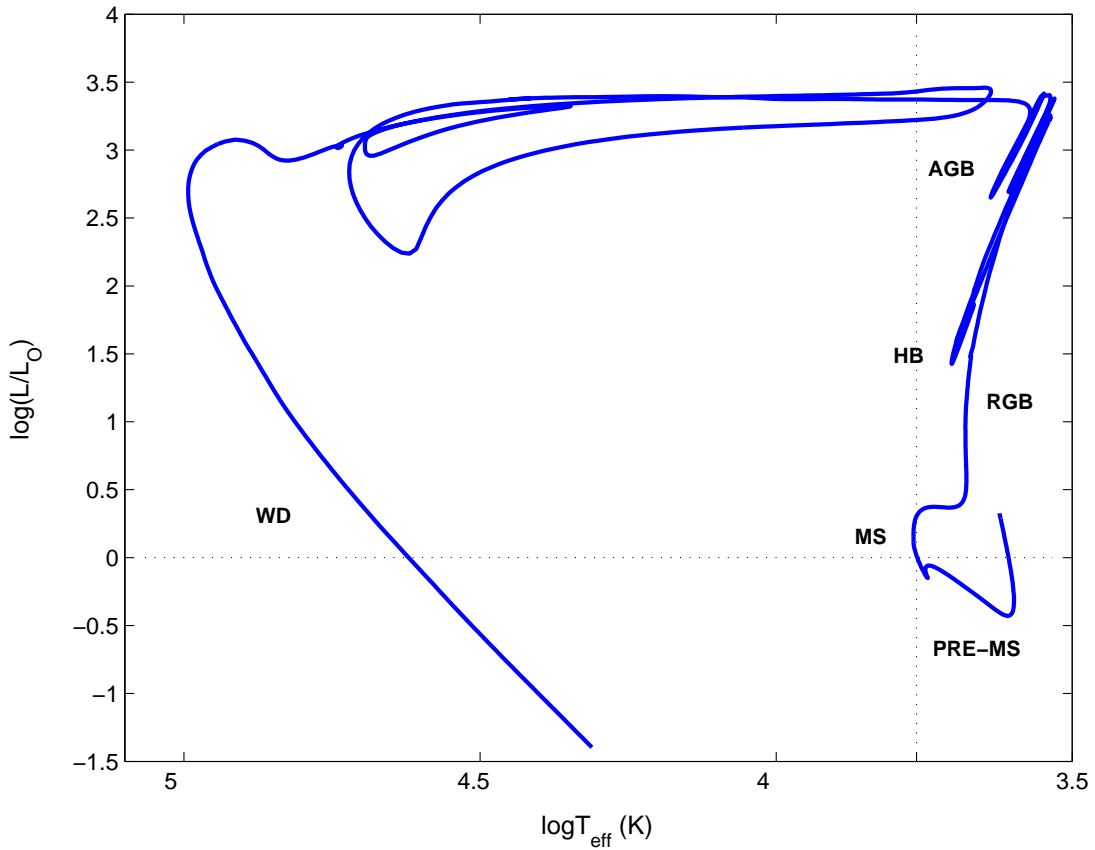}
\end{center}
\caption{Solar model HRD - A complete evolutionary track as obtained for $1 \Msun,\ Y=0.29,\ Z=0.018$ and mixing-length parameter $\alpha=2.5$.}
\label{fig:solar}
\end{figure}

Fig.~\ref{fig:solar} shows the evolutionary track in the HRD, where the various phases are marked:
from pre-MS, through MS, RGB and core HeF, settling into stable core He burning, continuing
through AGB and thermal pulses up to the last He shell
flashes---where a strong flash occurs, followed by a weaker one---and ending with a cooling
$0.55\,\Msun$ CO-WD.
The durations of the MS, RGB and HB stages are $10\ {\rm Gyr},\ 1.5\ {\rm Gyr\ and\ } 78\ {\rm Myr}$,
respectively. The maximum radius and luminosity---attained on the AGB after some $11.7$ Gyr
of evolution (from ZAMS)---are $1.46\times 10^2\ \Rsun$ and $2.81\times 10^3\ \Lsun$, respectively.
The maximum temperature throughout the evolution, $2.09\times10^8$~K, is attained off-center,
at the tip of the AGB.
We terminated the calculation with a final CO-WD of radius $R_{WD}=2.13\times10^{-2}\ \Rsun$, a central pressure
$p_{c,WD}=6.94\times10^{22}\,{\rm dyn}\,{\rm cm}^{-2}$, a central density
$\rho_{c,WD}=1.84\times 10^6\,{\rm gr}\,{\rm cm}^{-3}$ and a core temperature of $\sim 75$ million K.

\subsection{The Effect of Metallicity} \label{metal}

\begin{figure}
\begin{center}
\includegraphics[scale=.8]{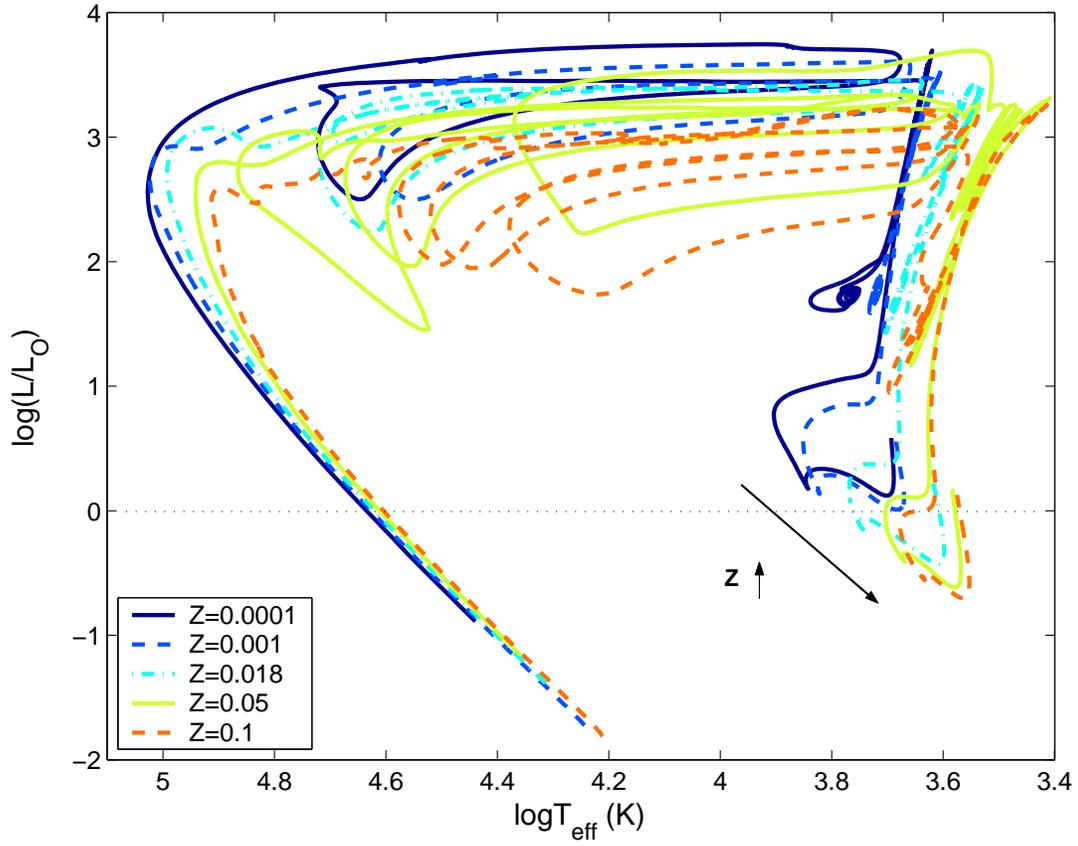}
\end{center}
\caption{Complete tracks on HRD for various metallicities $Z=0.0001\ to\ 0.1$ for $1 \Msun$.
MS effective temperatures and luminosities decrease with increasing metallicity;
consequently - durations of MS increase (by a factor of over $3$ from the lowest value of $Z$ to the highest).} 
\label{fig:metal}
\end{figure}

The effect of metallicity on stellar evolution is illustrated by a series of calculations for a model of solar mass
and $(Z,Y)$ values of (0.0001,0.24), (0.001,0.24), (0.018,0.29), (0.05,0.30) and (0.1,0.30), other physical and numerical
parameters remaining fixed. The results are presented in Fig.~\ref{fig:metal} by complete, continuous tracks in the H-R diagram.
We note that an increase in metallicity has a similar effect to a decrease in the initial stellar mass: luminosities are lower and
the durations of evolutionary phases are longer. For example, the MS phase lasts up to over 3 times longer, when $Z$ increases
from $10^{-4}$ to $0.1$.
This result is mostly the consequence of the dependence of opacity on composition;
at a lower metallicity, the opacity decreases, the star is able to radiate away its energy with greater efficiency,
the stellar luminosity is therefore higher and timescales are correspondingly shorter.

Apart from the apparent shift of the evolutionary tracks in the H-R diagram, and the different timescales,
metallicity also affects the final masses. For $M_i=1\ \Msun$, a final mass of $0.57\ \Msun$ was obtained for the lowest
metallicity ($Z=0.0001$), and $0.52\ \Msun$ for the highest one ($Z=0.1$), as compared with $0.55\ \Msun$,
obtained for solar metallicity---an overall spread of almost 10\%.

\subsection{Canonical Evolution Sequences} \label{canon}

\begin{figure}
\centering
\includegraphics[scale=.8]{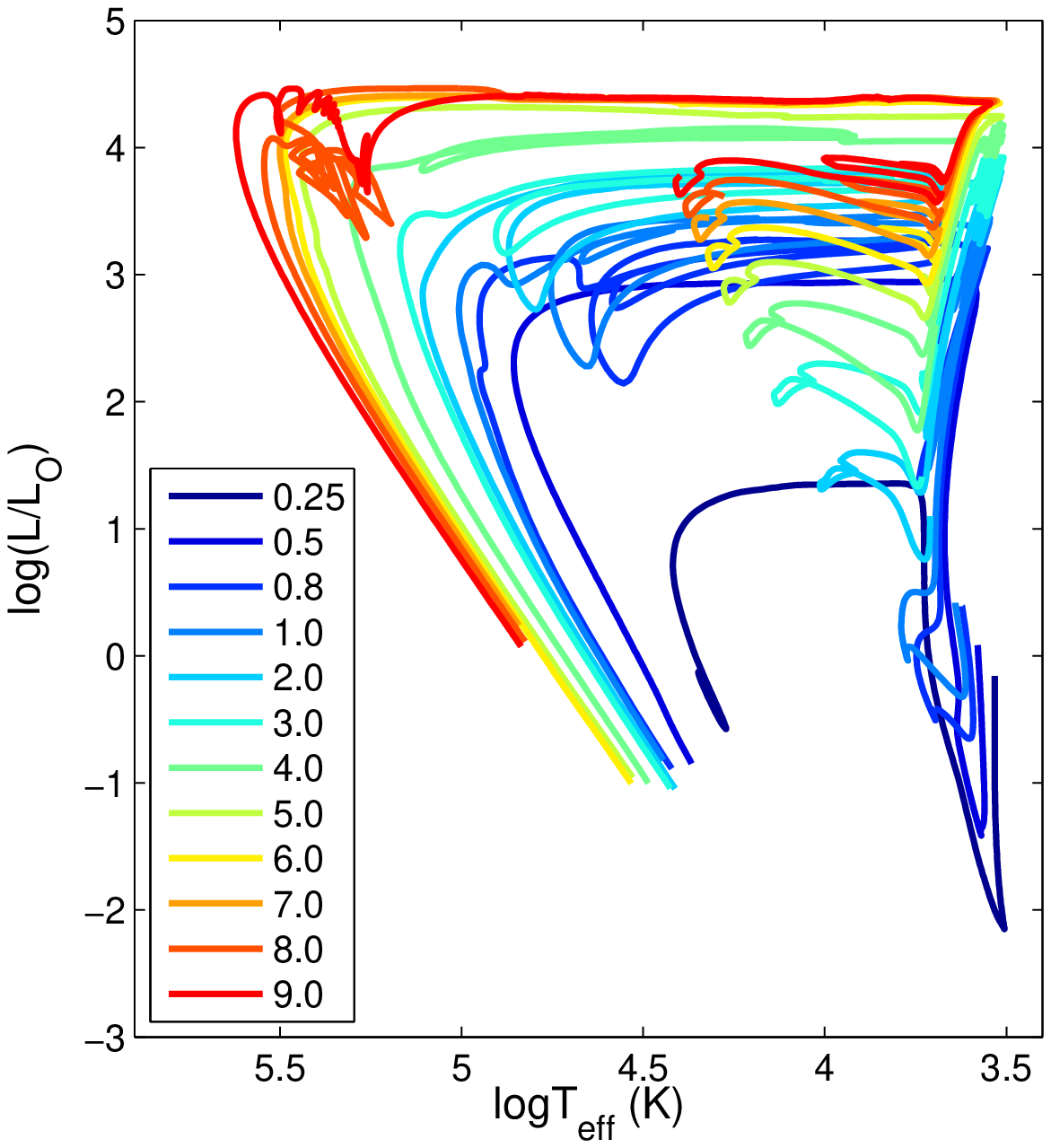}
\includegraphics[scale=.8]{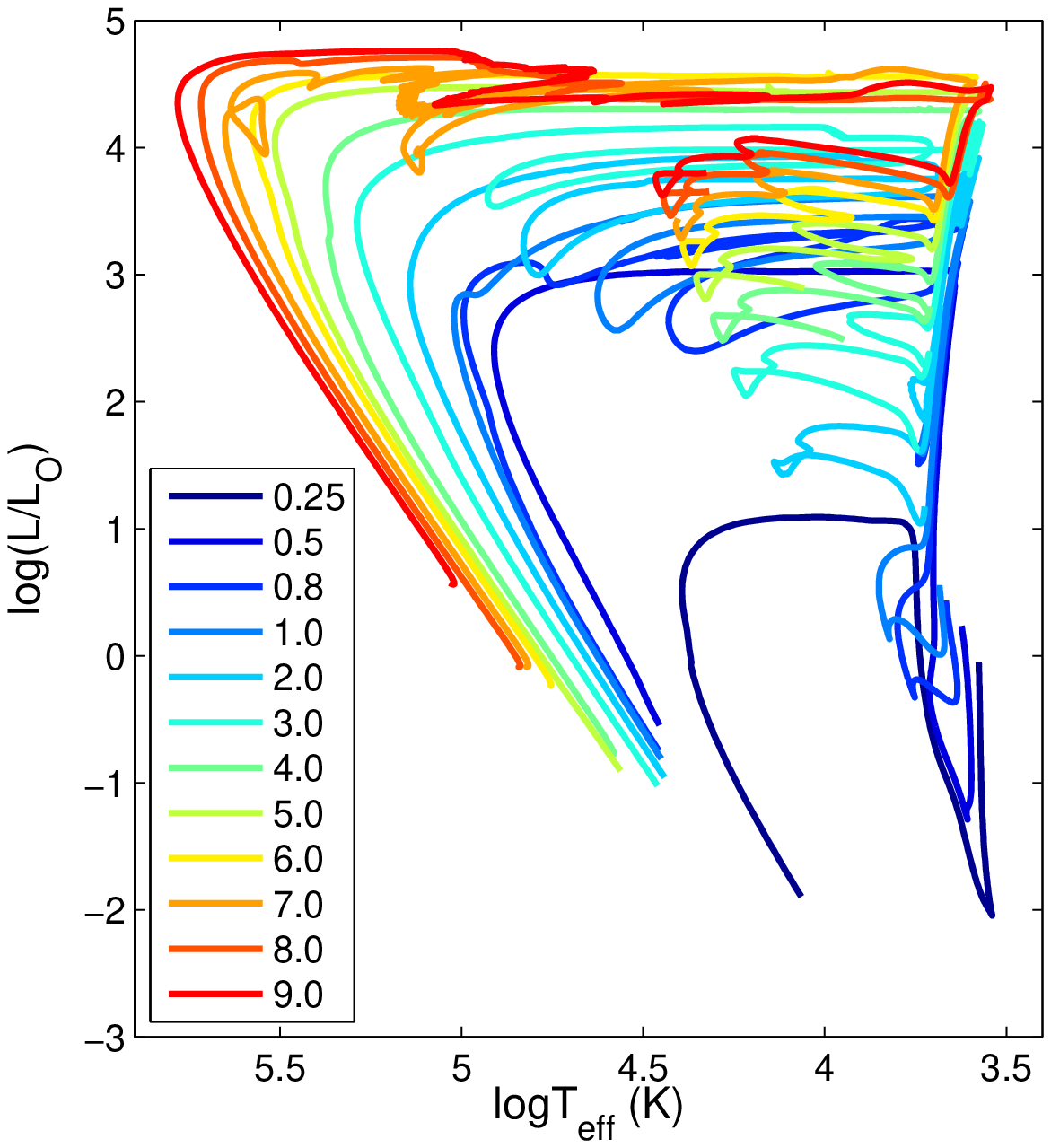}
\caption{Complete tracks on HRD---Pre-MS to cooling WD---for initial (ZAMS) masses in the range $0.25$ to $9.0\ \Msun$.
Top: population I models ($Z=0.01,\ Y=0.28$).
Bottom: population II models ($Z=0.001,\ Y=0.24$).}
\label{fig:hrds}
\end{figure}

We consider Population I (Pop.I) and Population II (Pop.II) stars, adopting metallicities of
$Z=0.01$ and $Z=0.001$, respectively, and initial masses in the range $0.25-9\,\Msun$, leading to
cooling WDs. The complete evolutionary tracks are shown in the two panels of Fig.~\ref{fig:hrds}.
Timescales and the final WD masses and composition are given in the accompanying Table~1. 
It should be noted that the MS and RGB durations as shown in the table depend on the definition of
the MS-turnoff point and beginning of the RGB, which involves some arbitrariness.
The criterion we use for the MS turnoff is as follows: let $t_1$ be the time when $X_c$ has decreased
below $10^{-6}$; let $x_1=\log T_{eff}(t_1)$ and $y_1=\log L(t_1)$. The turnoff time $t_2$ is the earliest
time for which the distance between the points [$x_2=\log T_{eff}(t_2)$ , $y_2=\log L(t_2)$] and [$x_1$ , $y_1$] in the
[$\log T_{eff}$ , $\log L$] plane exceeds 0.1. Similar criteria are used for other transitions between
evolutionary stages.
Time scales depend strongly on composition, especially on $Z$, decreasing with decreasing $Z$. Given differences
in composition adopted in different studies, as well as differences in criteria defining evolutionary stages,
a precise comparison between models is difficult to achieve. Nevertheless, we find excellent agreement, for
example, between our low-mass Pop.II models and corresponding ones calculated by others:
for the 0.8~$\Msun$ and 1~$\Msun$ models, we find $\tau_{MS}=1.44\times10^{10}$~yr and $6.02\times10^9$~yr, respectively
(see Table~1), while for the same masses and metallicity, \citet{1996A&AS..115..339C} find $1.51\times10^{10}$~yr 
and $6.06\times10^9$~yr, and \citet{1999A&AS..135..405C}, after modifying the input physics, find
$1.43\times10^{10}$~yr and $6.85\times10^9$~yr.   
For Pop.I models, the spread in initial $Z$ is larger, yet our results are compatible with those obtained
by \citet{2007ApJS..172..649S} for a grid of stelar models with $Z=0.019$.

The evolutionary tracks end with a cooling WD. A He-WD is obtained for the lower initial masses,
$0.25 \lesssim M_i\lesssim 0.50\ \Msun$.
The transition to a CO-WD occurs between $0.50$ and $0.80\ \Msun$,
and the heavier ONeMg-dominated WDs are obtained for initial masses higher than $\sim 8\ \Msun$
(the transition mass being higher for the Pop.I stars).
It should be noted, however, that especially for the Pop.I stars, the transition mass
for obtaining a CO-WD rather than a He-WD is strongly dependent on the mass-loss rate assumed.
For example, for an initial mass of $0.80\ \Msun$, slightly increasing the mass-loss rate may result either
in a He-WD, when the threshold for core helium burning is not reached,
or, in an Extreme Horizontal Branch (EHB) star, when a `delayed' core HeF takes place.
The production of such hot (blue) HB stars for relatively low initial masses (from around $0.80$ to slightly over $1\ \Msun$)
and for a range of metallicities will be addressed in a subsequent paper.

For both populations, a violent ignition of helium takes place in the core (but usually off-center, because of neutrino cooling)
at the tip of the first giant branch for masses in the range $0.80 - 2\ \Msun$. This is the well-known core HeF.
The transition between HeF and quiet He ignition occurs at an initial mass between $2$ and $3\ \Msun$,
depending mainly on composition and mass-loss rate. During the flash, the peak nuclear energy generation rate is in
the range $5\times 10^7 \lesssim L_{nuc,max} \lesssim 5\times 10^9\ \Lsun$,
decreasing with increasing initial mass, due to a corresponding decrease in the degree of electron degeneracy
of the core material. It is worth noting that the luminosity of the star during the flash is unaffected by what
is taking place in the core, despite the huge nuclear luminosity, which surpasses the luminosity obtained at any
evolutionary stage.
The overall duration of the flash (when $L_{nuc}$ is in excess of, say, $10^5\ \Lsun$) is of the order of a few years.
We note that during this stage time steps are automatically reduced down to days, then hours and minutes.
Once the flash is over, it will take some extra $10^3$ to $10^5$ years before the star settles into stable core He burning,
the HB phase.

\begin{figure}
\centering
\includegraphics[scale=.8]{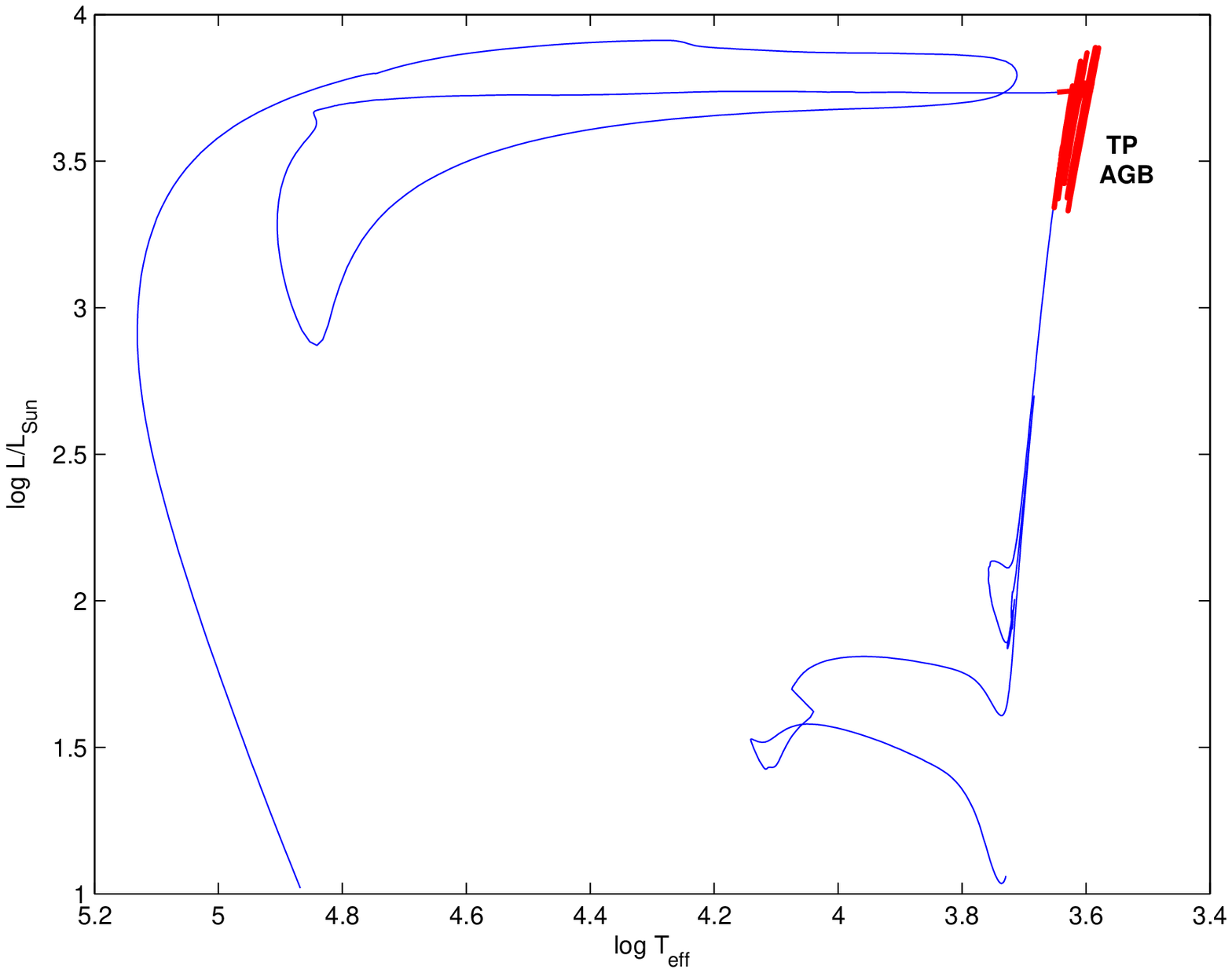}
\includegraphics[scale=.8]{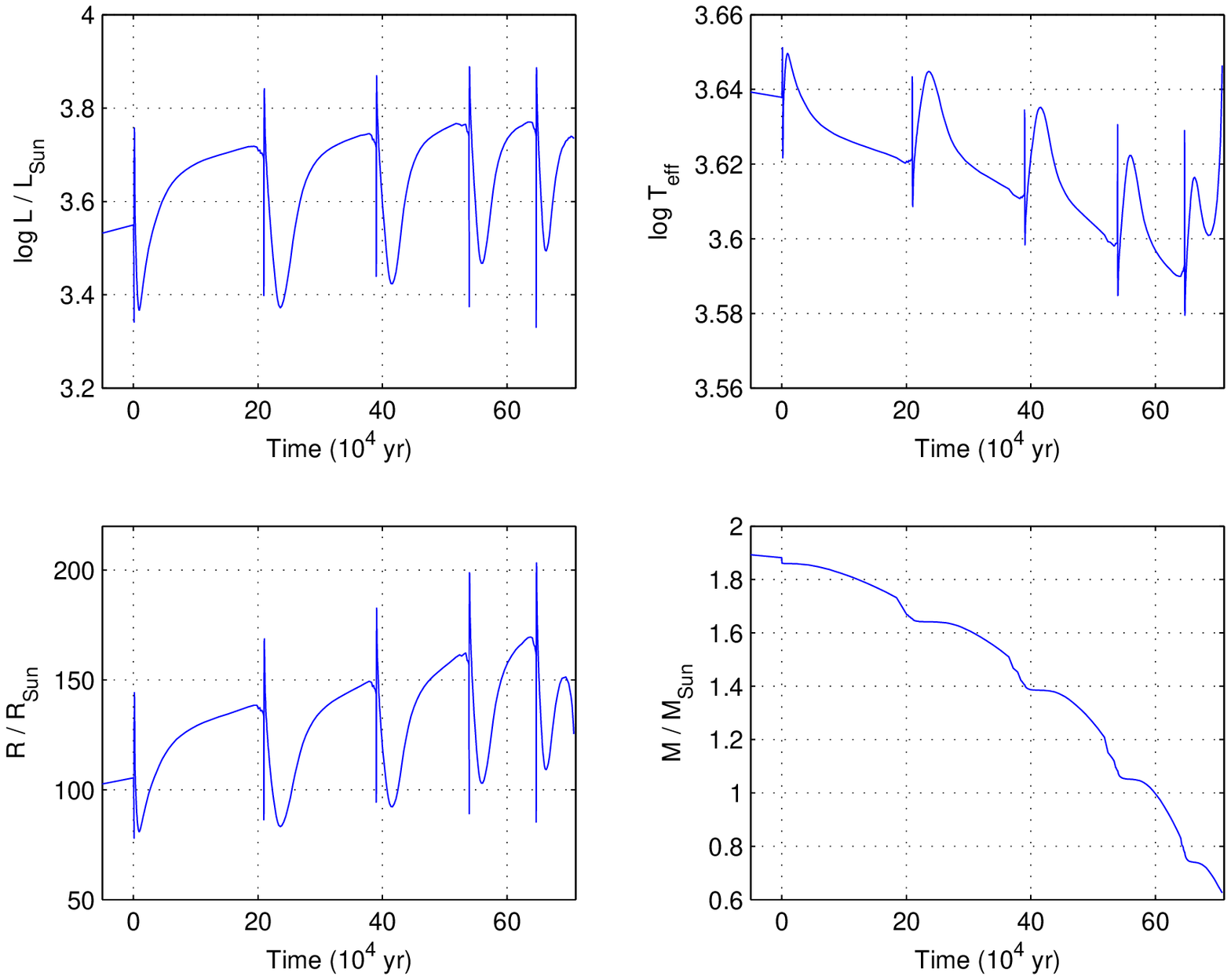}
\caption{Thermal pulses during TP-AGB for the $2\ \Msun$ Pop.II ($Z=0.001,\ Y=0.24$) model.
Top: complete track on HRD; plotted in thick red is the TP-AGB phase, for which the bottom panels are plotted.
Bottom: Evolution of various characteristics during the thermal pulses phase.}
\label{fig:tpagb}
\end{figure}

The well-known thermal pulses that arise as a result of the double shell-burning instability,
are clearly seen in the evolutionary sequences during the final stages of the AGB.
Fig.~\ref{fig:tpagb} shows a typical example for a Pop.II, $2\ \Msun$ model.
The thermal pulses in this example span about $7\times 10^5$~yr, and clear trends are evident,
such as the monotonic decrease in effective temperatures with advancing pulses,
along with an increase in the radial extension of the photosphere, which
reflect the asymptotic evolution towards the redder tip of the AGB.
Also evident is the fact that the bulk of mass-loss takes place precisely during this short phase, with
the mass dropping from $1.90$ to $0.63\ \Msun$ - almost its final value.
The mass of the H-depleted core increases during this phase from $0.58$ to $0.62\ \Msun$;
the mass of the inner He-depleted core increases from $0.48$ to $0.54\ \Msun$.
Since the He profile is not as steep as the H profile, the mass of the He-depleted core is a matter
of definition: here 'He-depleted' means $Y<10^{-6}\,$. Taking the core boundary at the mid-point of the He profile
yields a final He-depleted core mass of $0.60\Msun$.
We should note that the total number of pulses in each evolutionary sequence is largely determined by the
mass-loss law adopted.

\subsection{Mass-Loss Laws and Initial-Final Mass Relationship (IFMR)} \label{ifmr}

\begin{figure}
\begin{center}
\includegraphics[scale=.8]{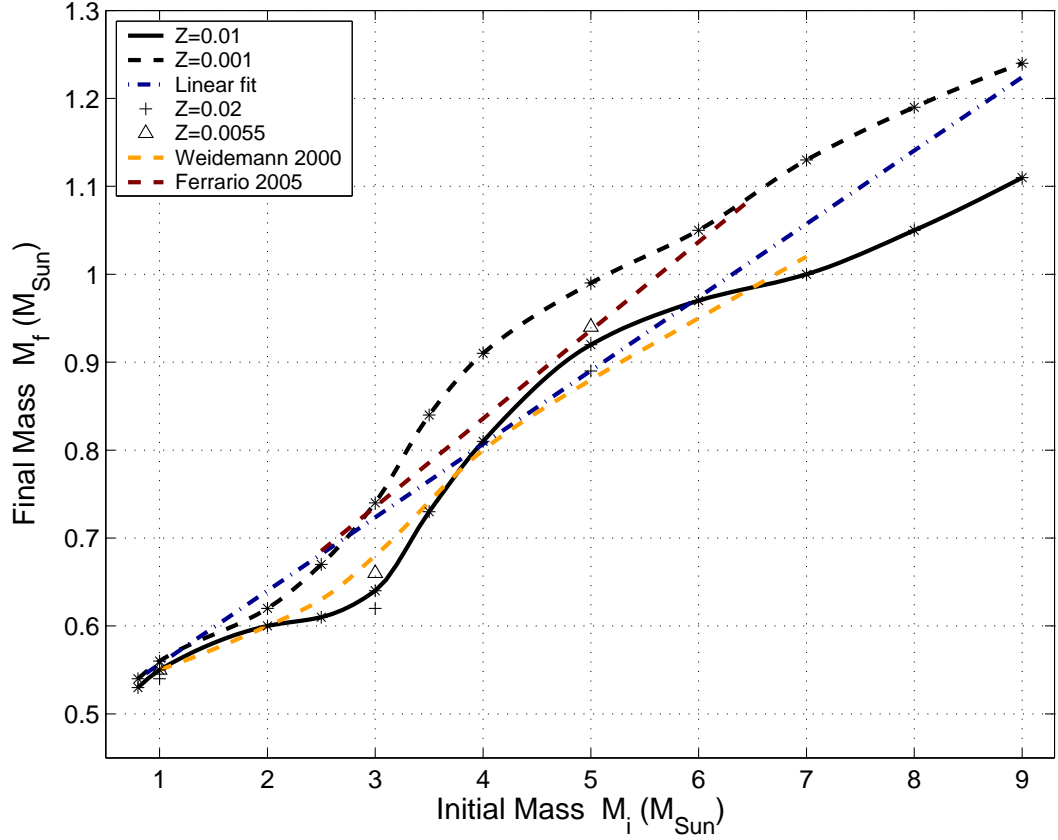}
\end{center}
\caption{IFMR - Final vs. initial masses as obtained for both our Pop.I (solid black)
and Pop.II (dashed black) evolutionary sequences,
for initial masses in the range $0.8$ to $9\ \Msun$.
The solid blue line is a linear fit to all values (Pop.I and II).
We show for comparison the revised \citet{2000A&A...363..647W} semi-empirical relationship ($M_i$ in the range $1$ to $7\ \Msun$),
as well as the empirical linear relation by \citet{2005MNRAS.361.1131F} ($M_i$ in the range $2.5$ to $6.5\ \Msun$).
See text for details.}
\label{fig:ifmr}
\end{figure}

Using the complete evolutionary tracks for the mass range of $0.8$ to $9\ \Msun$,
for both populations $Z=0.01$ and $Z=0.001$, as listed in Table~1, 
we obtain a theoretical IFMR, displayed in Fig.~\ref{fig:ifmr} (solid and dashed black lines).
We increased the number of points by adding results
for masses of $2.5$ and $3.5\ \Msun$, and for $Z=0.02$ and $Z=0.005$ and masses of $1,3$ and $5\ \Msun$
(marked in Fig.~\ref{fig:ifmr} by different symbols). Similar relationships have been recently
computed by \citet{2007arXiv0710.2397M} and by \citet{2008A&A...477..213C}, the latter including earlier results 
obtained by \citet{1999ApJ...524..226D}. A different and independent source for such a relationship is provided by 
observations (e.g., \citealt{2000A&A...363..647W}), mainly of star clusters, which lead to empirical or semi-empirical linear relations, 
such as \citet{2005MNRAS.361.1131F} (based on open-cluster data for the range $2.5-6.5\ \Msun$) and others that will be mentioned below. 

The curves obtained here show that the IFMR may be divided into three regions with different slopes:
1. A moderate slope  for $M_i\lesssim 3\ \Msun$, which
coincides with the tabulated results of \citet{2000A&A...363..647W} plotted in Fig.~\ref{fig:ifmr}.
2. A steeper slope for $3\lesssim M_i \lesssim 4\ \Msun$.
3. Again, a more gradual increase until the top end.
We note that the 'new relation' as displayed in Fig.~2 of \citet{1995LIACo..32..441H}, meant to fit only the best determined stars of the Hyades
and Pleiades clusters, has a very similar shape to our curves, only shifted upwards from our Pop.I curve by about $0.05\ \Msun$.

The dependence on metallicity is apparent from the divergence of the two curves in Fig.~\ref{fig:ifmr},
in agreement with the conclusions of e.g. \citet{2007arXiv0710.2397M} or the \citet{1999ApJ...524..226D} curves as plotted in Fig.~5 of \citet{2008A&A...477..213C}.
The effect of metallicity is negligible for $M_i\lesssim2\ \Msun$ (in agreement with e.g. \citet{2008A&A...477..213C}),
but it increases towards higher initial masses: the curves diverge by $\gtrsim0.13\ \Msun$ at the top end $M_i\gtrsim7\ \Msun$.
\citealt{2007arXiv0710.2397M} reach a difference of up to $0.4\ \Msun$ in the final masses derived from different metallicities,
their study covering a broad metallicity range: Z in between $0.0001$ and $0.1$. They also notice a minimum of the IFMR for $Z=0.04$.

Various semi-empirical linear fits have been derived over the last decade.
A few examples are:

\citet{2005MNRAS.361.1131F} (based on open-cluster data for the range $2.5-6.5\ \Msun$;
claiming that the IFMR can be modelled by a mean relationship about which there exists some intrinsic scatter,
and that they `cannot justify the use of any but a linear relationship to model the cluster data'):
\begin{equation} \label{eq:fer05}
     M_f=(0.10038\pm0.00518)M_i+0.43443\pm0.01467
\end{equation}

\citet{2006MNRAS.369..383D} (a linear fit to some $27$ WDs, members of clusters such as the Hyades, Praesepe, M35, NGC2516 and the Pleiades,
over initial-mass range of $2.7-6\ \Msun$):
\begin{equation} \label{eq:dob06}
     M_f=(0.133\pm0.015)M_i+0.289\pm0.051
\end{equation}

\citet{2007ASPC..372...85W} (claiming that the IFMR is both linear and without any metallicity dependence):
\begin{equation} \label{eq:wil07}
     M_f=(0.132\pm0.017)M_i+0.33\pm0.07
\end{equation}

Although the relations obtained, as shown in Fig.~\ref{fig:ifmr}, are quite far from linear,
the closest linear fit that we can suggest, without using any artificial anchoring, is
\begin{equation} \label{eq:ifmr_linear}
     M_f=0.08343*M_i+0.47321
\end{equation}
which falls slightly above the upper (Pop.II) curve around the lower initial masses ($1.5-2.5\ \Msun$),
and below the lower (Pop.I) curve for higher intermediate masses, around $5\ \Msun$.
This fit is very similar to the linear fit of \citet{2005MNRAS.361.1131F} (shown in Fig.~\ref{fig:ifmr}),
although the latter is limited to the range $2.5$ to $6.5\ \Msun$.

Clearly, the relation obtained represents the set of parameters assumed, mostly those related to the mass-loss recipe.
The value of $\eta_{\rm Reim}$ used here was linearly increased from 0.4 at $0.8\ \Msun$ to 3.0 at $9\ \Msun$.
A preliminary comparison that we performed, keeping all parameters fixed and changing only mass-loss laws,
indeed showed some differences in the final WD masses, with a spread of less than $10\%$.
More precisely, for our solar model parameters (see \S \ref{canon}), setting $\eta_{Rei}=0.6,\ R_{thresh}=50$,
the derived final WD masses were all in the range $0.53-0.57\ \Msun $
(or between $0.51-0.56$ for slightly higher mass-loss rates obtained by using $\eta_{Rei}=1.0,\ R_{thresh}=10$).
Performing the same comparison for $3\ \Msun\ (Z=0.01)$, but using $\eta_{Rei}=2.0$, we found final WD masses to be 
in the range $0.61-0.67\ \Msun $.

\subsection{Massive Stars}  \label{massive}

We now briefly consider Pop.I massive stars of initial masses in the range $16-64\ \Msun$, typically, SN progenitors. Since 
nucleosynthesis calculations are limited in our code, we cannot follow the evolution all the way to the collapse of an iron 
core. However, we come quite close to it. These massive stars go through advanced nuclear burning stages, until a core composed
of the end-product of our nuclear reactions network is obtained. Core masses range monotonically from 2.4$\Msun$ for the 
64$\Msun$ initial mass and 1.7$\Msun$ for the 16$\Msun$ initial mass. The core is enveloped by layers of different composition,
the outermost being predominantly helium. Envelope masses depend strongly on the mass loss law assumed.

\begin{figure}
\begin{center}
\includegraphics[scale=.8]{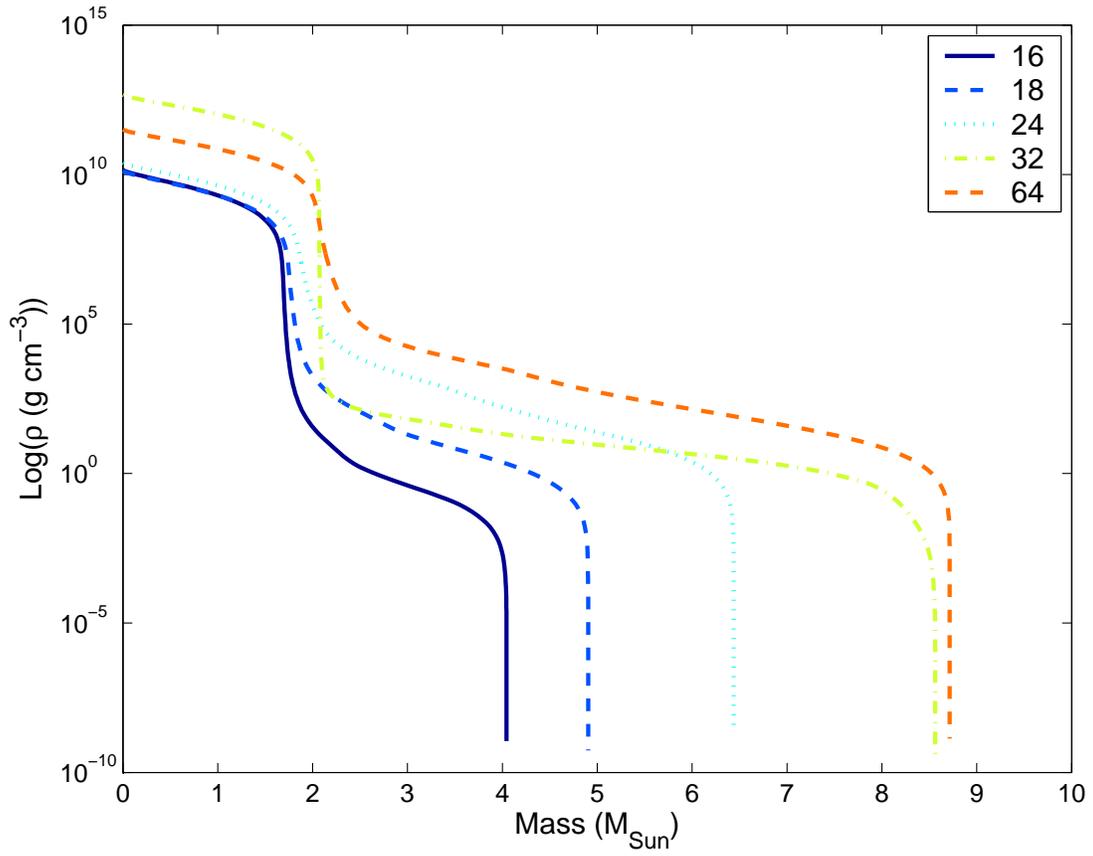}
\end{center}
\caption{Density profiles in massive stars (Pop.I) with collapsed cores (legend shows initial masses).}
\label{fig:rhoprof}
\end{figure}

The core contracts, becoming degenerate and unstable, since
its mass exceeds the Chandrasekhar limit. As contraction accelerates, temperatures rise to a few $10^{10}$~K, where electron-positron pairs are
created, which enhances the instability, lowering the adiabatic exponent. Pair production replaces iron photodisintegration as
the mechanism leading to core collapse. Density profiles throughout the stars are shown in Fig.~\ref{fig:rhoprof}.
The code crashes when the collapse approaches free-fall, with the adiabatic exponent very close to 4/3 throughout the core.
Since this point is somewhat arbitrary, the curves representing stars of different initial masses do not exhibit a perfectly 
regular (monotonic) behaviour; this is sometimes the case
for evolutionary tracks or characteristics of massive stars in the late stages 
(\citealt{1996snai.book.....A}, \citealt{2008ApJ...673.1014U}), resulting from the complexity 
of the processes taking part in them, and the related parameters and thresholds.
We do not claim that these calculations shed light on pre-supernova evolution; rather, we mention them here as an example of the robustness 
of the code, which is capable of dealing with complex processes under critical conditions without failing.

\begin{figure}
\begin{center}
\includegraphics[scale=.8]{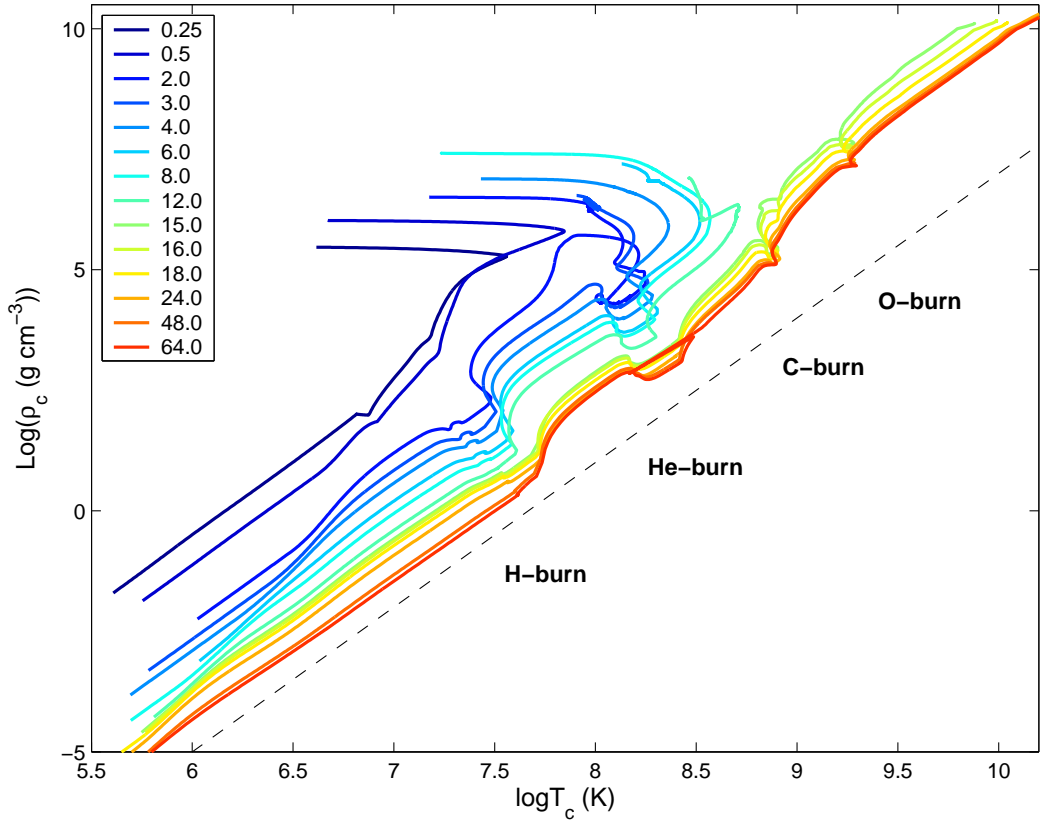}
\end{center}
\caption{Evolution of the central stellar density and temperature for Pop.I 
($Z=0.018$) models in the range $0.25\ -\ 64\ \Msun$.
Dotted line has a slope of 3 (as obtained for the $\log\rho_c-\log T_c$ relation of hydrostatic equilibrium under ideal gas law).
Nuclear burning phases are marked along the tracks.}
\label{fig:rhoctc}
\end{figure}

Finally, adding the results obtained for lower masses of Pop.I, described in Section~\ref{canon}, we show in Fig.~\ref{fig:rhoctc} 
evolutionary tracks of the stellar central points in the $(\log T,\log\rho)$ plane, exhibiting
the branching off between stars that end their lives as WDs, and stars that go
through advanced nuclear burning stages, ending their lives in dynamic core collapse.

\subsection{Non-Canonical Evolution} \label{noncanon}

The term `non-canonical' refers to stars of unusual internal structure and composition.
Such configurations may result from stellar mergers, where the merging stars may be MS stars,
giants, compact stars or any combination of different types.
Stellar mergers are probably the progenitors of blue straggler stars (BSS),
found to exist in environments of high stellar density,
such as globular clusters or the cores of open clusters.

As already mentioned, the main reason for developing the evolution code presented here was the need
for an efficient and fast tool that could be integrated into
the MODEST (MOdelling DEnse STellar systems) collaboration,
combining dynamical N-body calculations with hydrodynamics---the colliding or merging of stars---and
stellar evolution, for the simulating of dense stellar environments.
Whereas for normal stars, it is possible to construct and tabulate pre-computed evolutionary tracks for
the use of MODEST calculations, merger products, having completely unpredictable configurations, must be evolved
{\it in situ}.

A non-canonical initial model will be the product of a hydrodynamic merger calculation, usually by smoothed
particle hydrodynamics (SPH) methods. The first step in adapting such a model to quasi-static
stellar evolution calculations is to obtain a hydrostatically relaxed configuration. This is achieved by
applying the quasi-dynamic method of \citet{1967ApJ...150..131R}. Instead of eqs. \eqref{eq:evol1}--\eqref{eq:evol2},
consider the equations

\begin{equation} \label{eq:qd1}
        \frac{1}{\rho}=\frac{\partial}{\partial m}\frac{4\pi}{3} r^3  ,
\end{equation}
\begin{equation} \label{eq:qd2}
        \frac{\partial r}{\partial\tau}=-4\pi r^2\frac{\partial p(\rho,s,Y)}{\partial m}-\frac{Gm}{r^2}  ,
\end{equation}

\noindent
where $r(m,\tau)$ is regarded as a function of the mass coordinate $m$ and the quasi-time $\tau\,$, and
$p(\rho,s,Y)$ is determined by the EOS. The quasi-time has no physical meaning: its purpose is provide
asymptotically (i.e. for $\tau\rightarrow\infty$) a hydrostatic solution. Equation \eqref{eq:qd2} is
called quasi-dynamic because the correct dynamic equation would have $\partial^2 r/\partial t^2$---with
$t$ the true time---on its left-hand side.

Let the boundary conditions be $r=0$ at the center, and $p=0$ at the surface.
For a given distribution of entropy $s(m)\,$, and of the number
fractions, collectively denoted by $Y(m)\,$, and an initial distribution of radii $r(m,0)\,$,
the foregoing equations are to be solved for $r(m,\tau)$ (and $\rho(m,\tau)\,$, and $p(m,\tau)\,$).

Since the entropy $s$ and the composition $Y$ are not varied, the (quasi) motion is adiabatic: $du=-pd(1/\rho)\,$.
Multiplying \eqref{eq:qd2} by $\partial r/\partial\tau$ and integrating over the mass of the star yields,
after an integration by parts,
\begin{equation} \label{eq:qd3}
\int_0^M\Bigl(\frac{\partial r}{\partial\tau}\Bigr)^2dm=-\frac{dE}{d\tau} ,
\end{equation}
where
\begin{equation} \label{eq:qd4}
  E=\int_0^M\Bigl( u-\frac{Gm}{r}\Bigr)\,dm
\end{equation}
is the total energy, internal and gravitational. Equation \eqref{eq:qd3} shows that the energy decreases with quasi-time.
If---for the given entropy and composition distributions---a minimum of $E$ exists, the solution of eqs.
\eqref{eq:qd1}--\eqref{eq:qd2} must lead to it, and the resulting structure, of stationary energy, will be hydrostatic.
If, on the other hand, a minimum of $E$ does not exist, the configuration is dynamically unstable: $E$ will then
decrease indefinitely.

Thus, the quasi-dynamic method either leads to a hydrostatic structure, or else detects dynamical instability.
It can be applied to any initial density distribution, even a uniform one. With the EOS
\begin{equation} \label{eq:qd5}
  p(\rho,s,Y)=K\rho^{1+\frac{1}{n}}
\end{equation}
it can be used to construct a polytrope (dynamically unstable when $n\geq 3$), which may serve as an initial 
`fully convective' protostellar model of uniform entropy and composition. Of course, `solution' of
\eqref{eq:qd1}--\eqref{eq:qd2} entails the replacement of the differential equations by implicit difference equations,
which are then solved by an iterative process \citep{1967ApJ...150..131R}.

As preliminary examples, we evolved merger products for three pairs of Pop.II ($Z=0.001$) low-mass parent stars.
The parent stars were evolved by our code from some pre-MS initial configuration, 
to an age when the more massive star of each pair was almost at terminal MS age (TAMS), 
the less massive star of the pair being, of course, at an earlier stage on the MS.
A pair of $0.85$ and $0.60\ \Msun$ parent stars was evolved for $11$ Gyr;
a pair of $1.00$ and $0.60\ \Msun$ for $6$ Gyr;
and finally, a pair of $1.40$ and $0.60\ \Msun$ for $1.5$ Gyr.
To calculate structures of the merger products for the above pairs of parent stars,
we used the MMAS (`make me a star', version 1.6) package of \cite{2002ApJ...568..939L},
which produces 1D models that approximate results of detailed SPH calculations.
We chose to perform head-on collisions (zero periastron separation), so that effects of rotation were absent.
Each resulting merger product was incorporated {\it as is} into our code, and upon obtaining a hydrostatically relaxed configuration
by the `quai-dynamic method' as explained above, calculation of the evolution was initiated. 

It might be worthwhile to note the difference between the way we treat the merger-product
and the way the non-canonical evolution is initiated by \citet{2008A&A...488.1007G}, \citet{2008A&A...488.1017G}.
As explained in these papers, what the authors did was to start from a ZAMS model of the correct mass, evolve
it until the central $X_H$ equalled that of the merger product and then evolve it further with a fictitious energy production
until its entropy profile equalled that of the merger product. This was done in steps, during which the composition was 
gradually adjusted to that of the merger product. This process resulted in a hydrostatic configuration that had 
the given mass and correct entropy and composition profiles.
In contrast, what we did was to make use of the merger product exactly as obtained by the collision calculation
and subject it to the quasi-dynamic method. 

Table~\ref{tab:noncanon} lists some details of the colliding stars and the resulting mergers:
$t_{col}$ is the time of collision (age to which the parent stars were evolved);
$M_{merger}$ is the mass of the merger product (slightly less than the sum of parent star masses,
because some mass was lost in the merger process); $Y_c$ is the central He mass-fraction,
$\tau_{MS}$ is the remaining MS lifetime of the merger-product, whereas $\tau_{MS,counter}$ is the MS duration of the canonical counterpart -
a normal (`canonical') star of initial mass equal to that of the merger-product.
The central He mass fraction generally depends on the stages to which the parent stars have been evolved - 
how close to TAMS was the more massive parent star, and correspondingly, 
how much hydrogen did the less massive star of the pair managed to burn during its limited MS evolution.
It should be noted, for instance, that the MS duration of the $1.88\ \Msun$ merger-product exceeds that of the lower-mass $1.48\ \Msun$ merger-product;
this is due to the greater amount of central hydrogen in the more massive merger-product.

\begin{figure}
\begin{center}
{\includegraphics[scale=.4]{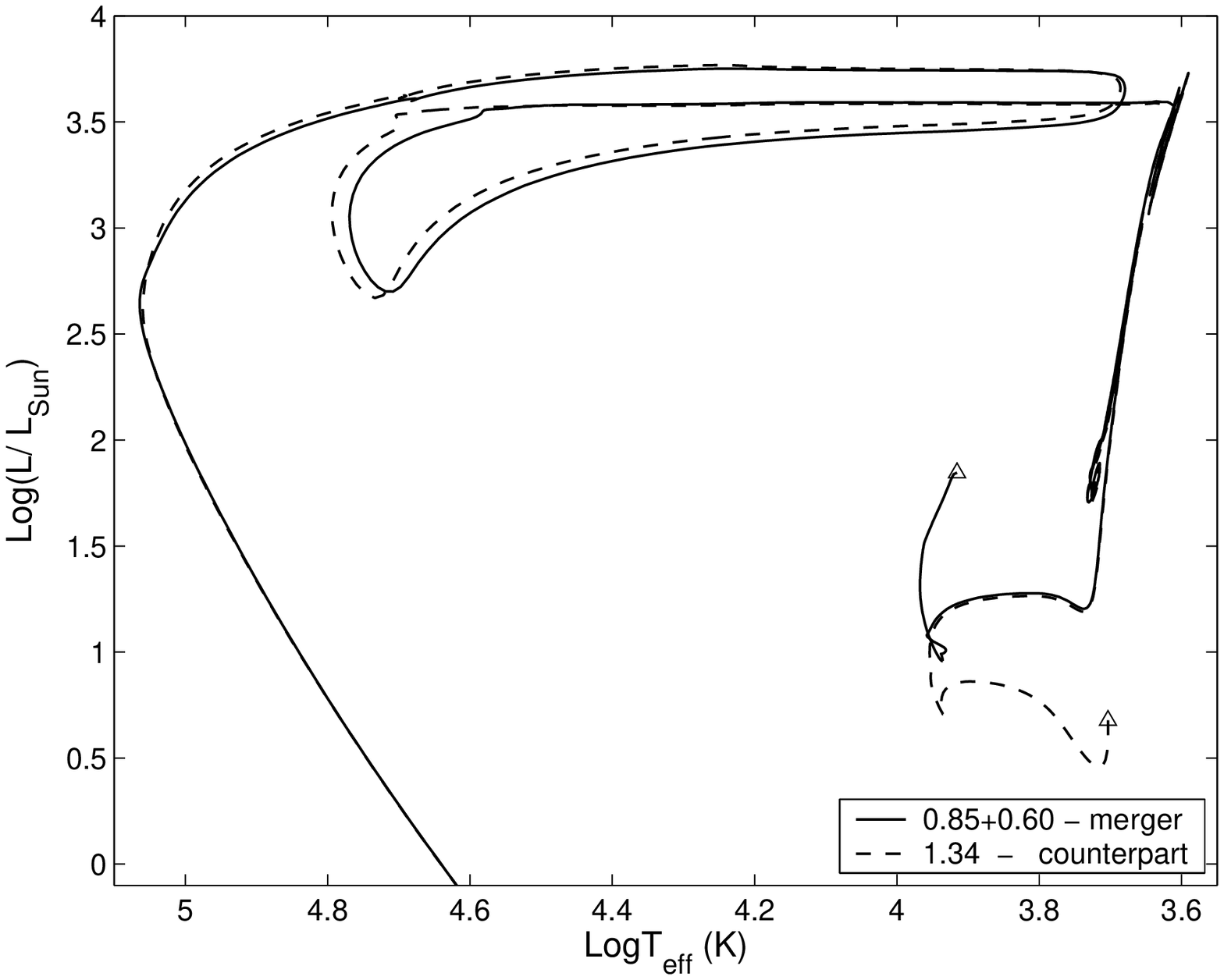}
 \includegraphics[scale=.4]{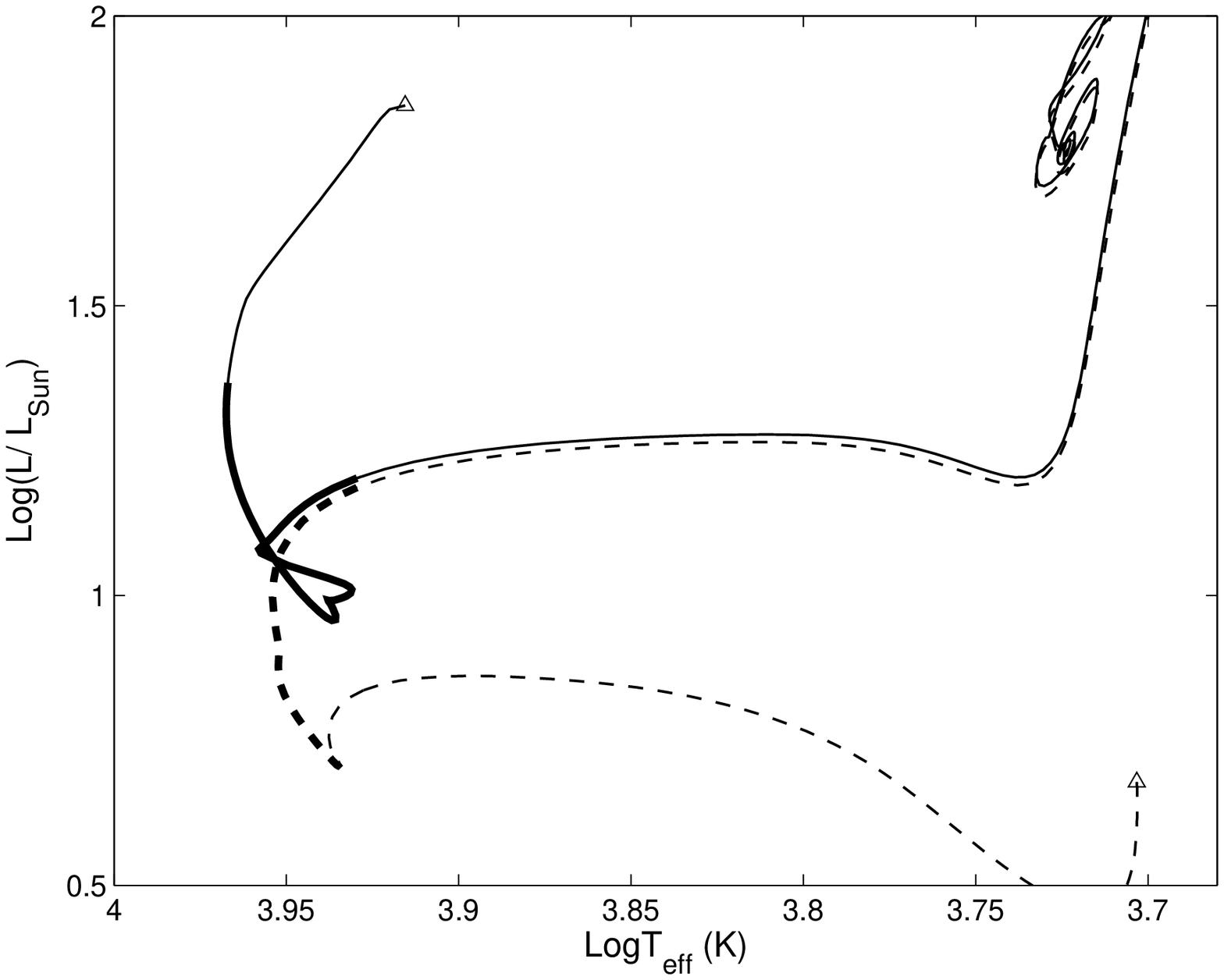}}
{\includegraphics[scale=.4]{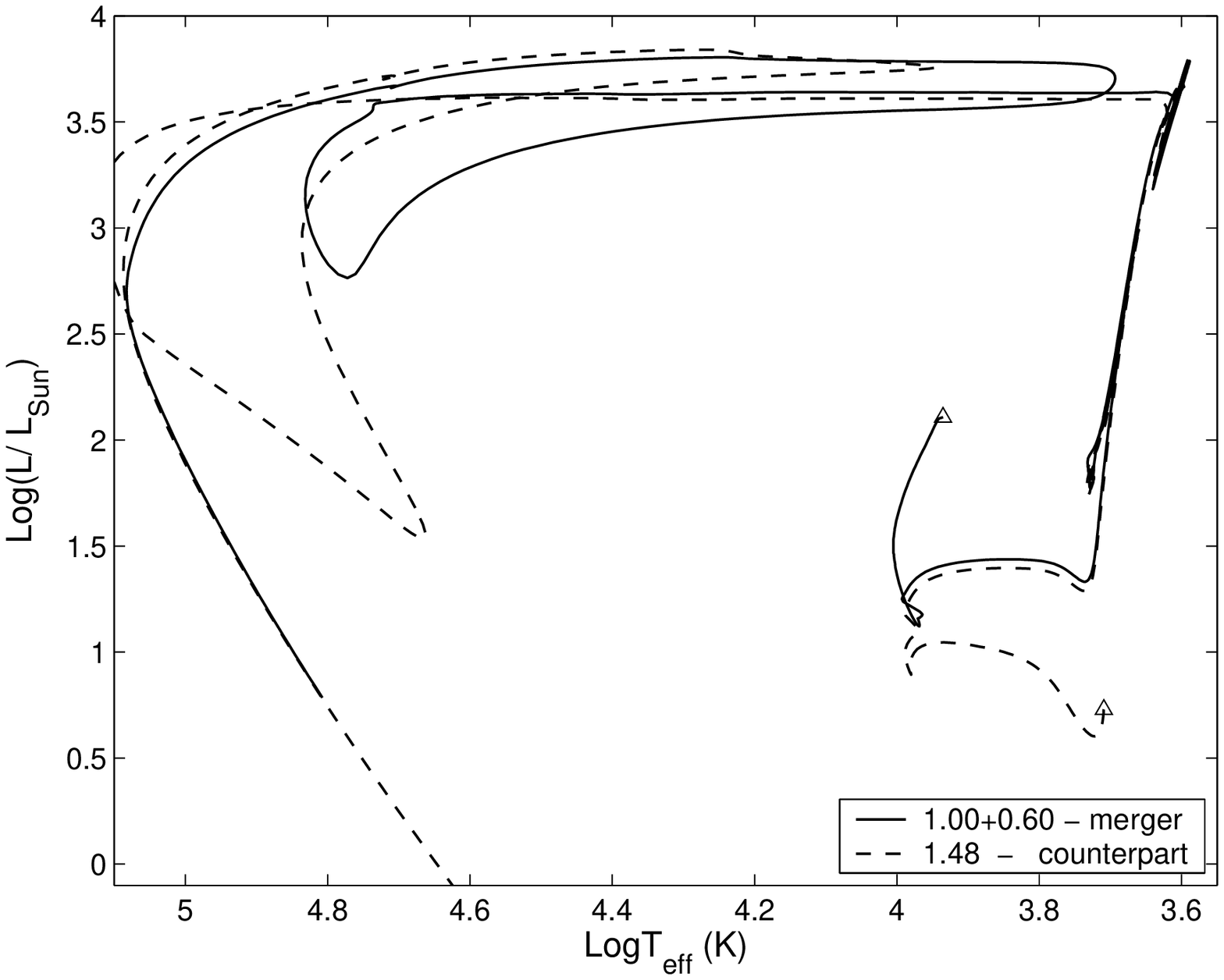}
 \includegraphics[scale=.4]{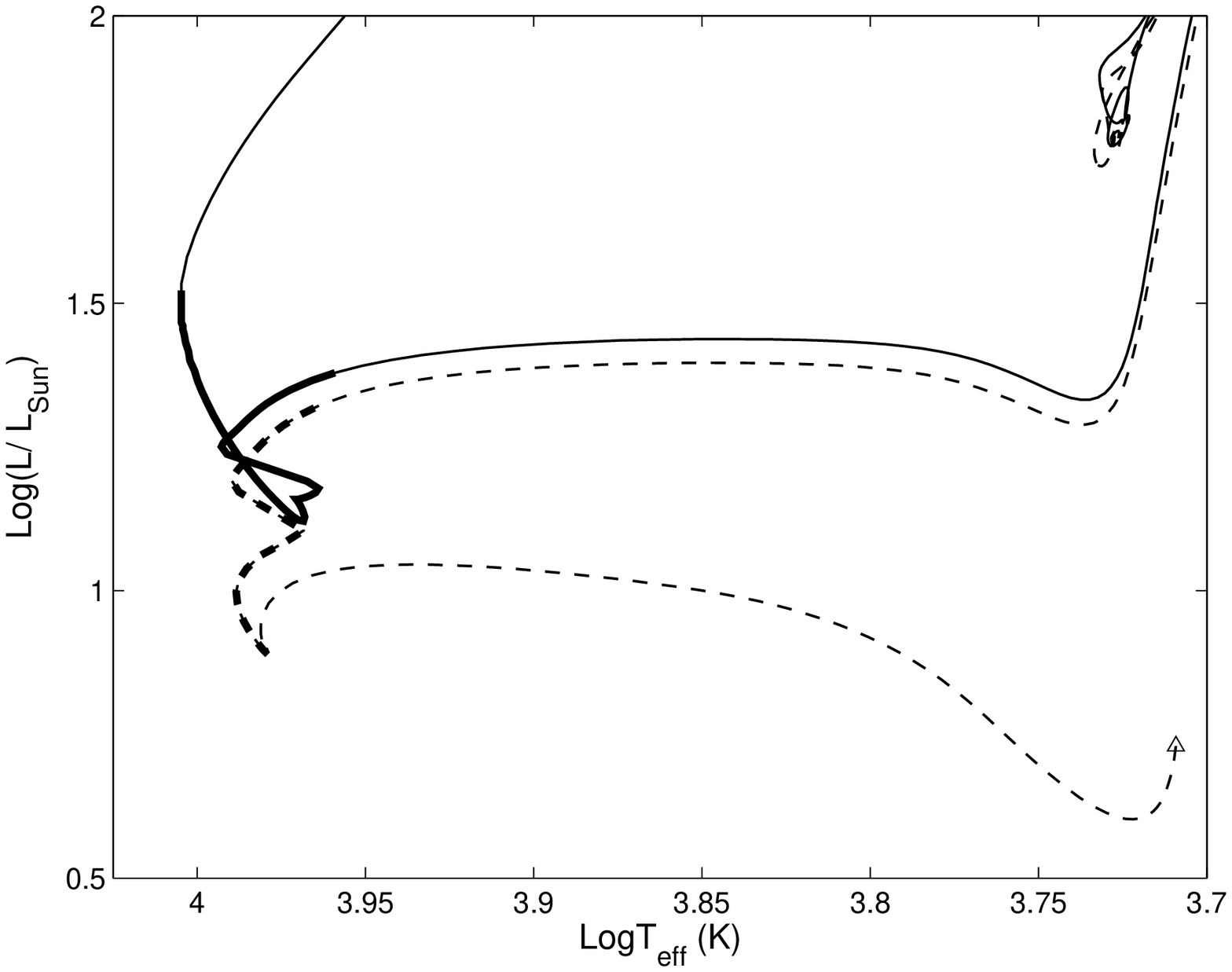}}
{\includegraphics[scale=.4]{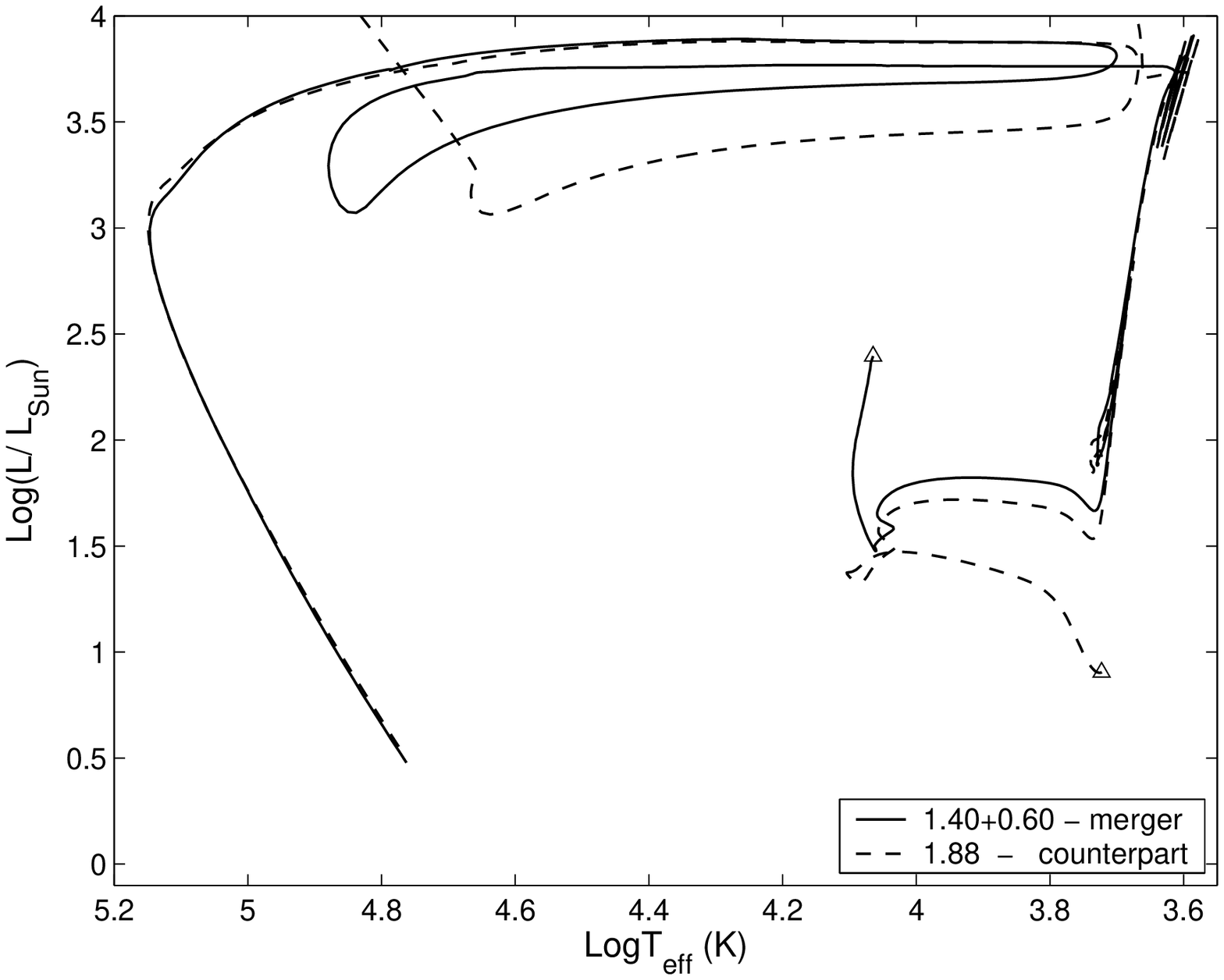}
 \includegraphics[scale=.4]{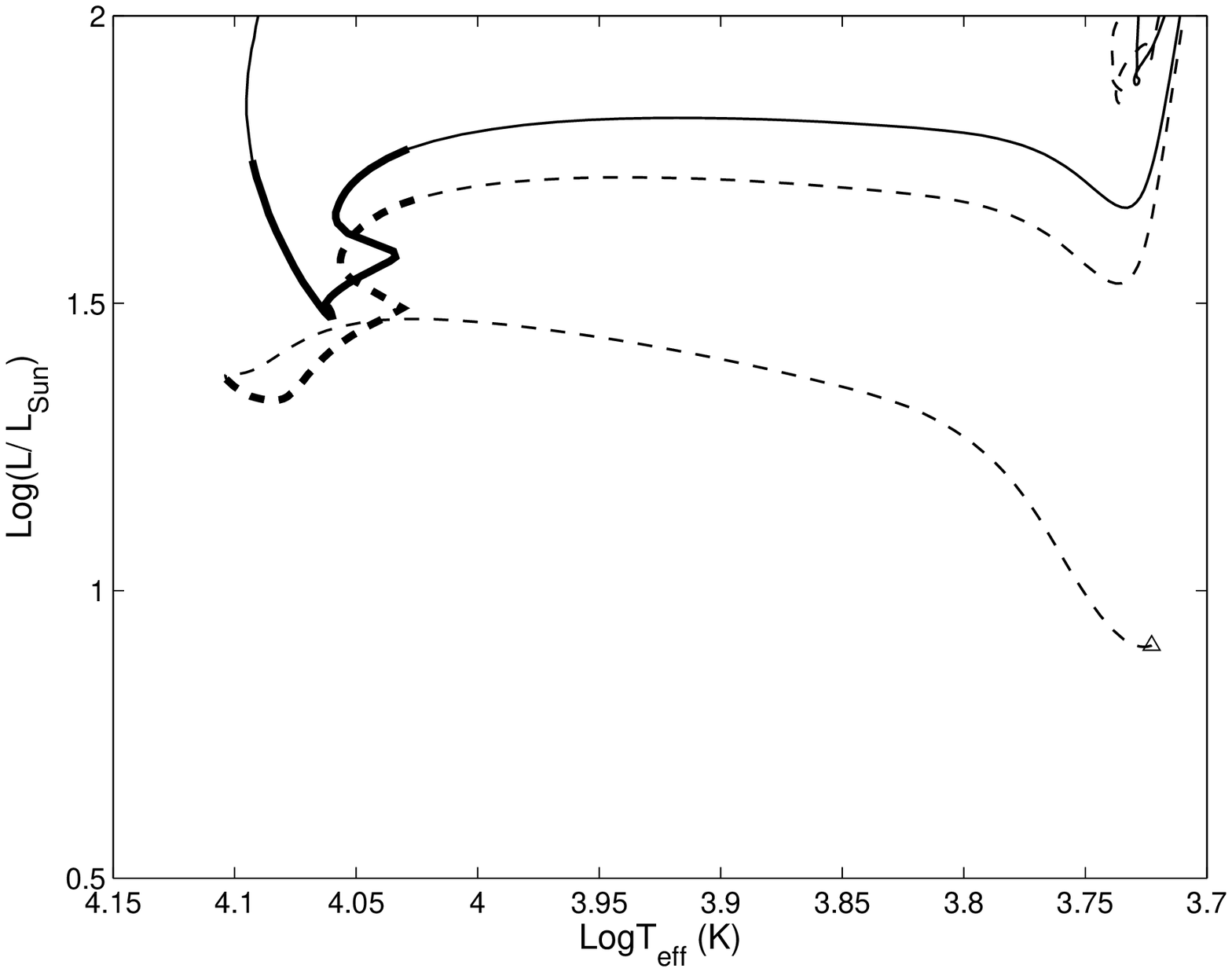}}
\end{center}
\caption{Non-canonical evolution - evolutionary tracks on HRD of the three merger products (solid) 
of low-mass MS parent stars,
with comparison to their canonical counterparts - `normal' initial configurations of equal mass (dashed).
Triangles denote starting points for both the canonical and non-canonical evolutionary tracks.
Panels on the right are a blow-up of the (MS,early-RGB) regions;
thicker solid and dashed sections denote the extent of MS evolution phase.
}
\label{fig:noncanon_hrd}
\end{figure}

Fig~\ref{fig:noncanon_hrd} shows evolutionary tracks on HRD of the three merger products (solid lines) 
($0.85+0.60,\ 1.00+0.60,\ 1.40+0.60$ - top to bottom), while
dashed lines represent evolutionary tracks of the canonical counterparts.
The non-canonical models, possessing excess thermal energy right after the merging process, 
all begin by gravitational contraction before settling on the MS, where they spend the time required 
for burning the remaining central hydrogen.
It is only during the MS and early-RGB phases that the non-canonical track differs from that of the canonical one.
The non-canonical track is shifted slightly upwards (to higher luminosity);
the shift is growing with increasing mass (as is clearly apparent in the blow-up panels on the right).
Except for the insignificant differences in the shape of the last shell flash while traversing the HRD from the AGB tip to the
cooling WD curve, the tracks almost exactly overlap from RGB onwards.

\begin{figure}
\begin{center}
{\includegraphics[scale=.45]{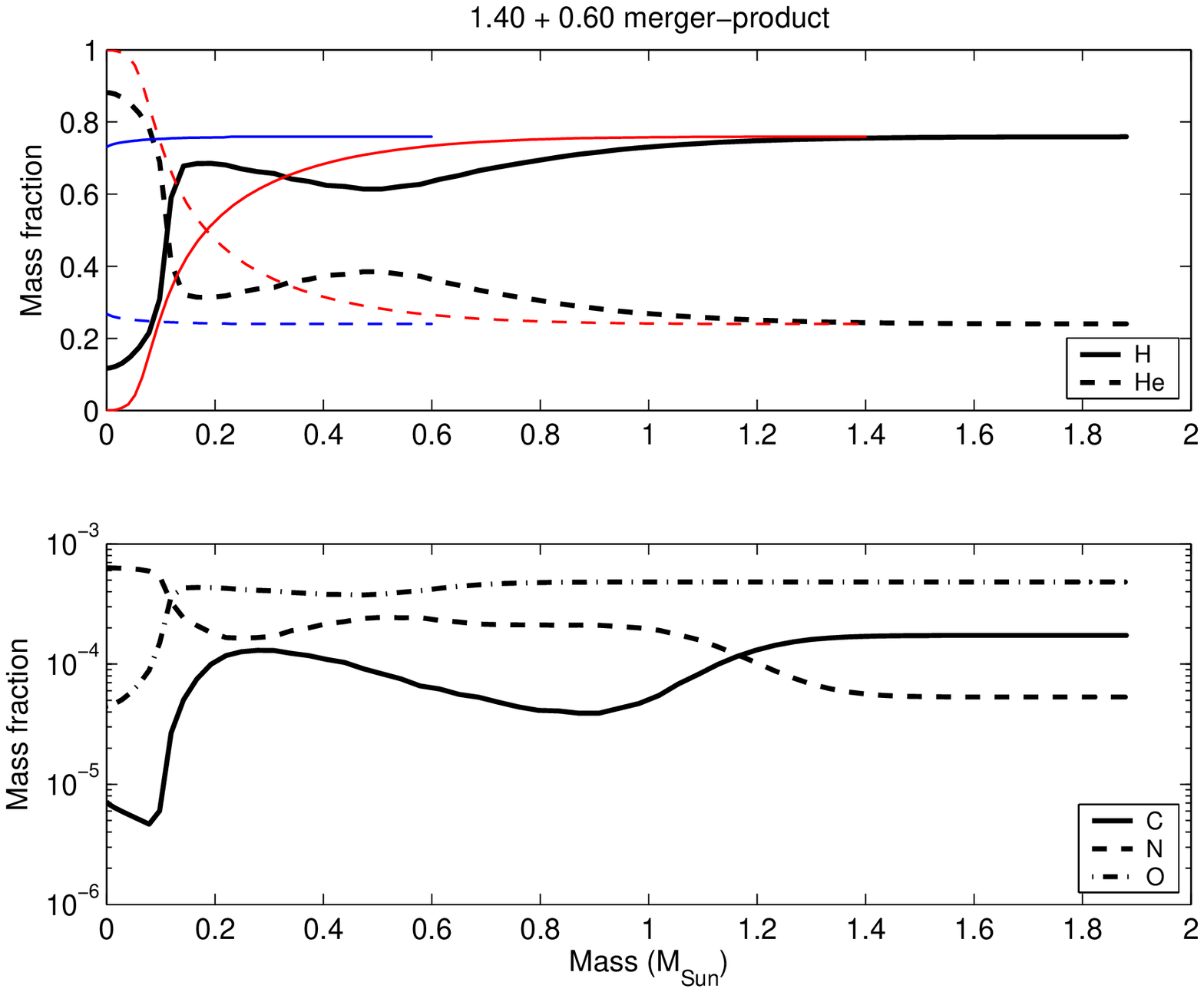}
 \includegraphics[scale=.45]{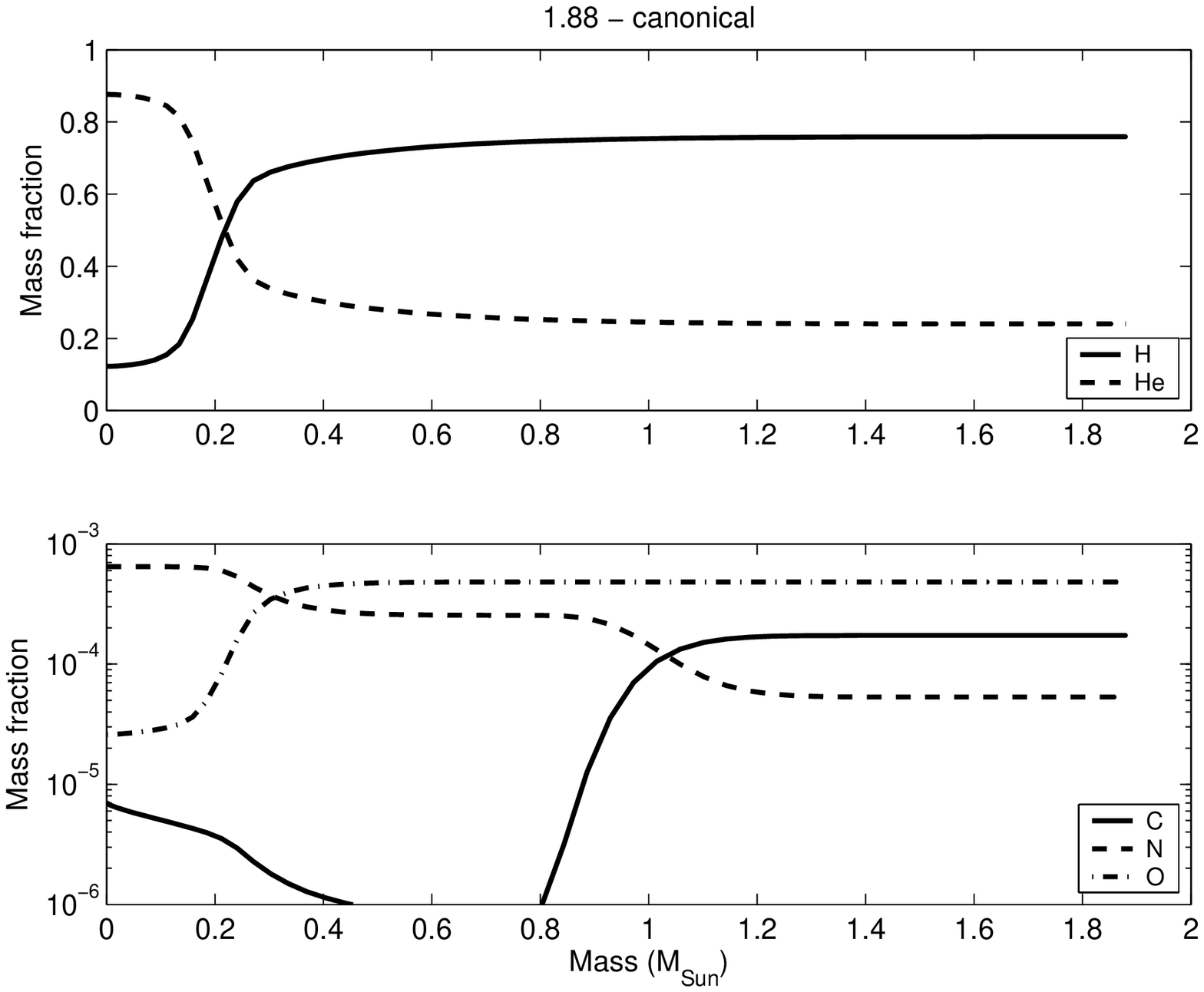}}
\end{center}
\caption{Left: Composition profiles (top: H,He; bottom: C,N,O) of the $1.40+0.60\ \Msun$  merger-product,
as obtained from the Make Me A Star ver 1.6 package for a head-on collision (see text), 
right after the configuration has been hydrostatically relaxed by our code -- ready to be evolved.
Shown in thin lines at the top panel are the H and He profiles of the $1.40\ \Msun$ (red) and $0.60\ \Msun$ (blue) parent stars,
evolved to an age of $1.5$ Gyr.
Right: similar profiles for the canonical counterpart -- a `normal' $1.88\ \Msun$ star -- 
when its central He mass fraction equals that of the merger product.}
\label{fig:noncanon_prof}
\end{figure}

Fig.~\ref{fig:noncanon_prof} shows, as an example, composition profiles of the $1.40+0.60\ \Msun$ merger,
with comparison to the $1.88\ \Msun$ canonical counterpart at the point when the latter's $Y_c$ equals 
that of the initial state of the merger product.
In the top panel the H and He profiles of the merger-product are plotted together with those of the 
$1.40$ and $0.60$ parent stars.
Central hydrogen is almost completely depleted for the more massive parent star, which is very close to its TAMS;
the low-mass parent star, at early stages of its MS evolution, still has a large fraction of hydrogen.

Already a decade ago \cite{1997ApJ...487..290S} began investigating evolutionary scenarios of collisionally merged stars,
with the aim of examining possible formation channels and properties of blue straggler stars in globular clusters.
They present results of evolutionary calculations for seven head-on collisions.
Among their results, we find for instance a MS duration of $3.74\times10^8$~yr for their $0.80+0.60\ \Msun$ merger;
although details of the collision, including abundances, might not be exactly comparable, this result seems to be in very 
good agreement with our derived MS duration of $3.75\times10^8$~yr for our $0.85+0.60\ \Msun$ similar merger.

As mentioned, more extensive evolutionary calculations for collision products 
have recently been performed by \citet{2008A&A...488.1007G}, \citet{2008A&A...488.1017G}.
Nowadays, several procedures for performing calculations of stellar collisions,
such as the mentioned MMAS by \cite{2002ApJ...568..939L}
or MMAMS (`make me a massive star') by \cite{2008MNRAS.383L...5G} are available.
As illustrated by the foregoing three examples, our code is able to import and initiate evolution for merger-products 
created by either of the above procedures.
In future, it will be interesting to study non-canonical evolution merger-products 
over a wider range of masses and initial compositions (outcomes of various combinations of the parent stars),
as well as mergers involving other types of stars, such as compact objects---the merging of WD-MS or WD-WD.

\begin{table} \label{tab:noncanon}
 \centering
  \caption{Collisions of low-mass MS parent stars - characteristics of the parent stars and merger-products.
           Masses are in solar units; MS durations are in years.}
  \begin{tabular}{llcccccc}
  \hline
$M_1$ & $M_2$ & $t_{col}\ (Gyr)$  &  & $M_{merger}$ & $Y_c$ & $\tau_{MS}$ & $\tau_{MS,counter}$ \\
\hline
0.85  & 0.60  & 11.0        & \vline & 1.34  & 0.96       & 3.75e8      & 2.05e9  \\
1.00  & 0.60  & 6.0         & \vline & 1.48  & 0.98       & 2.84e8      & 1.43e9  \\
1.40  & 0.60  & 1.5         & \vline & 1.88  & 0.88       & 2.91e8      & 6.10e8  \\
\hline
\end{tabular}
\end{table}

\section*{Summary}

We have developed a stellar evolution code that is capable of calculating full evolutionary tracks without interruption or
intervention. The implicit numerical scheme is based on simultaneous solution of the thermodynamic and composition equations on
an adaptive grid. Time steps are self-adjusting according to numerical as well as evolutionary time-scale criteria.
The code was applied to a large variety of examples: full evolutionary tracks for stars of a wide range of
masses and metallicities, and non-canonical stars obtained from stellar mergers.
We believe that these examples of stellar evolution calculations demonstrate the efficiency and rubustness of
our new code. We mention, in particular, the ability of the code to deal with the core He flash, thermal pulses, WD cooling, core collapse,
as well as non-canonical configurations. We thus expect it to be useful in extensive parameter studies---of both stellar
physics and initial properties of stellar models---as well as in simulations of stellar clusters.

\section*{Acknowledgments}

We are grateful to James Lombardi for providing us with his MMAS code.
We also wish to thank Onno Pols for a very careful reading of the original manuscript and numerous comments and
helpful suggestions. This work was supported in part by the Israel Science Foundation grant 388/07.

\vfill\eject

\bibliographystyle{mn2e}
\bibliography{mybibfile}

\vfill\eject

\begin{table} \label{tab:seq}
 \centering
  \caption{Timescales and final WD masses and compositions - Populations I and II canonical sequences}
  \begin{tabular}{lcccccccccc}
  \hline
      & & \multicolumn{4}{c}{Pop. I ($Z=0.01$)} & & \multicolumn{4}{c}{Pop. II ($Z=0.001$)}\\
$M_i$ & & $M_f$  & $\btau_{MS}$ & $\btau_{RGB}$ & WD   & & $M_f$   & $\btau_{MS}$ & $\btau_{RGB}$ & WD\\
\hline
0.25 & \vline & 0.25   & 8.70E+11    & 1.58E+10     & He    & \vline & 0.25    & 5.41E+11    & 1.77E+10     & He    \\
0.50 & \vline & 0.41   & 1.18E+11    & 6.38E+09     & He    & \vline & 0.46    & 9.10E+10    & 4.54E+09     & He    \\
0.80 & \vline & 0.53   & 1.98E+10    & 1.91E+09     & CO    & \vline & 0.54    & 1.44E+10    & 1.18E+09     & CO    \\
1.00 & \vline & 0.55   & 8.00E+09    & 1.24E+09     & CO    & \vline & 0.56    & 6.02E+09    & 7.51E+08     & CO    \\
2.00 & \vline & 0.60   & 7.43E+08    & 8.70E+07     & CO    & \vline & 0.62    & 5.04E+08    & 1.00E+08     & CO    \\
3.00 & \vline & 0.64   & 2.56E+08    & 2.75E+07     & CO    & \vline & 0.74    & 2.00E+08    & 2.18E+07     & CO    \\
4.00 & \vline & 0.80   & 1.25E+08    & 9.97E+06     & CO    & \vline & 0.92    & 1.10E+08    & 9.16E+06     & CO    \\
5.00 & \vline & 0.92   & 6.88E+07    & 5.80E+06     & CO    & \vline & 0.99    & 7.00E+07    & 4.84E+06     & CO    \\
6.00 & \vline & 0.97   & 4.52E+07    & 3.48E+06     & CO    & \vline & 1.05    & 4.91E+07    & 2.86E+06     & CO    \\
7.00 & \vline & 1.00   & 3.27E+07    & 2.08E+06     & CO    & \vline & 1.15    & 3.34E+07    & 2.15E+06     & CO    \\
8.00 & \vline & 1.05   & 2.44E+07    & 1.82E+06     & CO    & \vline & 1.20    & 2.60E+07    & 1.51E+06     & ONeMg \\
9.00 & \vline & 1.16   & 2.23E+07    & 1.11E+06     & ONeMg & \vline & 1.24    & 1.63E+07    & 1.26E+06     & ONeMg \\
\hline
\end{tabular}
\end{table}

\bsp

\label{lastpage}

\end{document}